\documentclass[review]{elsarticle}

\usepackage{lineno,hyperref}
\usepackage{xcolor}
\usepackage{amsmath,amssymb,amsfonts}
\usepackage{longtable}
\usepackage{bm}
\usepackage{hhline, colortbl,xcolor, multirow}
\usepackage{booktabs}

\journal{Information Fusion}

\begin{document}

\begin{frontmatter}

\title{Multimodal Audio-Visual Information Fusion using Canonical-Correlated Graph Neural Network for Energy-Efficient Speech Enhancement}

\author[1]{Leandro A. Passos}	
\author[2]{Jo\~{a}o Paulo Papa}
\author[5,6]{Javier Del Ser}
\author[3]{Amir Hussain}
\author[1,4]{Ahsan Adeel}

\address[1]{CMI Lab, School of Engineering and Informatics, University of  Wolverhampton \\
	Wolverhampton, England, United Kingdom.}
\address[2]{Department of Computing, S\~ao Paulo State University\\
	Bauru, S\~ao Paulo, Brazil.}
\address[5]{TECNALIA, Basque Research \& Technology Alliance (BRTA)\\
	Derio, Bizkaia, Spain.}
\address[6]{University of the Basque Country (UPV/EHU)\\
	Bilbao, Bizkaia, Spain.}
\address[3]{School of Computing, Edinburgh Napier University\\
	Edinburgh, Scotland, United Kingdom.}
\address[4]{deepCI.org, Edinburgh, Scotland, United Kingdom\\
	email:ahsan.adeel@deepci.org.}

\begin{abstract}

\begin{sloppypar}
This paper proposes a novel multimodal self-supervised architecture for energy-efficient audio-visual (AV) speech enhancement that integrates Graph Neural Networks with canonical correlation analysis (CCA-GNN). The proposed approach lays its foundations on a state-of-the-art CCA-GNN that learns representative embeddings by maximizing the correlation between pairs of augmented views of the same input while decorrelating disconnected features. The key idea of the conventional CCA-GNN involves discarding augmentation-variant information and preserving augmentation-invariant information while preventing capturing of redundant information. Our proposed AV CCA-GNN model deals with multimodal representation learning context. Specifically, our model improves contextual AV speech processing by maximizing canonical correlation from augmented views of the same channel and canonical correlation from audio and visual embeddings. In addition, it proposes a positional node encoding that considers a prior-frame sequence distance instead of a feature-space representation when computing the node's nearest neighbors, introducing temporal information in the embeddings through the neighborhood's connectivity. Experiments conducted on the benchmark ChiME3 dataset show that our proposed prior frame-based AV CCA-GNN ensures a better feature learning in the temporal context, leading to more energy-efficient speech reconstruction than state-of-the-art CCA-GNN and multilayer perceptron. 
\end{sloppypar}
\end{abstract}


\end{frontmatter}


\section{Introduction}
\label{sec:introduction}

Recent technological advances have empowered computers to reason and perform activities once attributed to human beings only, such as writing, speaking, and even helping with decisions like the best route on a trip or the most appropriate drugs for treating a specific disease. Nevertheless, such reasoning is limited to the domain upon which the algorithm is trained, i.e., the actions and decisions adopted by the algorithm are based on patterns somehow encoded in the dataset. This approach seems unnatural if considering the learning processes performed by the biological brain, in which stimuli are provided by a set of different sensors, e.g., vision and hearing, and this multimodal information is combined in such a way that redundant information is essential to reinforcing and improving noisy, ambiguous, and imperfect signals from distinct sources. 

Multimodal learning approaches are be beneficial to improve one modality feature representation~\cite{singh2022watmif,iqbal2022ff}. To illustrate the idea, consider, for instance, a boisterous speech. If the speaker only resorted his voice to convey its message, the comprehension of the subject may be considerably impaired by the noise. On the other hand, if the images are also available, it is possible to try to complement the corrupted information with insights provided by this secondary source, such as following the speaker's hands and body movements or trying to read his lips. 

Despite the availability of contextual information provided by different input signals, usually such tasks also rely on temporal information for reasoning. Revisiting the speech example, it is easy to infer that words being said at the present moment are probably strongly correlated with the last few pronounced words. Even though some works addressed the problem using recurrent networks~\cite{mai2021analyzing,li2020multimodal}, most of them perform target-driven supervised learning, which usually requires a considerable number of labeled samples for training.

In this context, unsupervised or self-supervised algorithms show themselves capable of extracting strongly correlated features, which are highly desired for two main reasons: (i) their representations are usually more general-purpose than target-driven features extracted with supervised algorithms, and (ii) they do not require labeled instances for training, which are usually limited and costly. Regarding the proposed approaches for correlated feature extraction, one can refer to energy- and mutual information-based methods, such as the Deep Graph Infomax~\cite{velivckovic2018deep}, which relies on mutual information maximization and graph neural networks (GNN) for leveraging information propagation in a graph.

With the advent of deep learning, GNNs emerged as an elegant solution to extract in-depth dependencies from such intricate relationships. Moreover, it also presents itself as a powerful alternative to convolutional neural networks suitable for datasets composed of non-imagery data. Despite the success obtained by~\cite{velivckovic2018deep}, Zhang et al.~\cite{zhang2021canonical} pointed towards a set of drawbacks in the model: (i) reliance on negative samples - corrupting the graph structure by selecting arbitrary negative examples may lead to large variance for stochastic gradients and slow training convergence; (ii) require a parameterized estimator to approximate mutual information between two views; and (iii) it contrasts node embeddings with graph embedding leading to a higher complexity. To tackle such problems, they propose the GNNs with Canonical Correlation Analysis (CCA-GNN), which aims at maximizing the correlation between two augmented views of the same input while decorrelating different dimensions of such views. Besides, Dwivedi et al.~\cite{dwivedi2021graph} exposed another shortcoming regarding GNNs message-passing mechanism, which builds node representation by aggregating feature space-based local neighborhood information and leads representations dependants on the local structure of the graph and proposed using positional encoding to solve the problem. 

\begin{figure*}[!ht]
  \centerline{
    \begin{tabular}{c}
	\includegraphics[width=\textwidth]{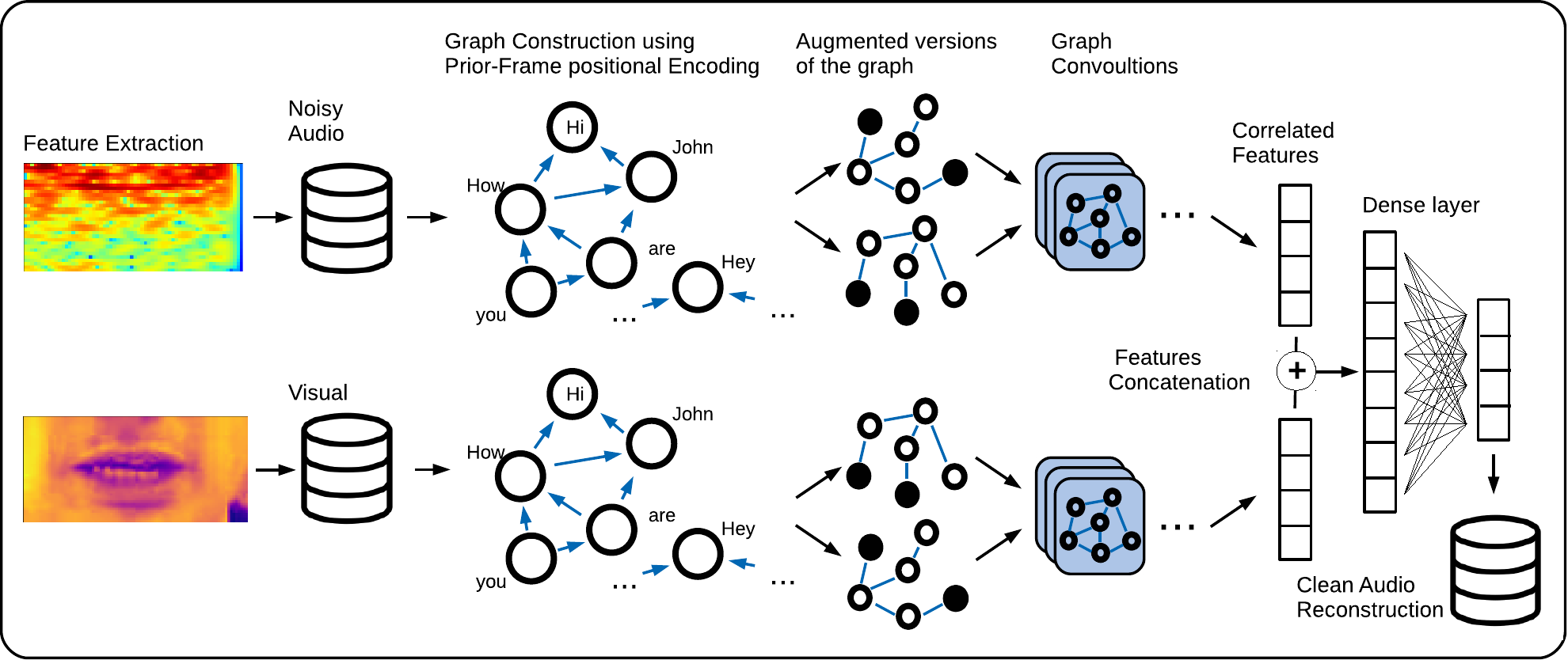} 
    \end{tabular}}
    \caption{Proposed pipeline. Features are extracted from the noisy audio input using Fourier transformation followed by logarithmic compression, while the visual inputs employ an encoder-decoder approach followed by Viola-Jones~\cite{viola2001rapid} for lip-regions identification. Both sets are converted into graphs using the proposed prior-frame-based positional encoding to maintain the temporal information. Such graphs are used to generate two augmented versions per epoch, which feed a set of graph convolutions. The model minimizes canonical correlation analysis of convolutions' output, aiming to generate a set of correlated features in a self-supervised fashion. Such features are finally used to feed a supervised dense layer responsible for reconstructing the clean audio signal.}
  \label{f.intro}
\end{figure*}

Therefore, this paper aims to redesign the GNNs with Canonical Correlation Analysis (CCA-GNN) to deal with the challenging context of multimodal representation learning. Specifically, it formulates a parallel CCA-GNN architecture for each input channel, i.e., audio and visual. The new audio-visual (AV) CCA-GNN model minimizes both the canonical correlation between the augmented samples of the same channel as well as between the augmented samples of the other mode. Additionally, it introduces graph modeling that considers a time-frame sequence distance positional encoding to compute the node's neighborhood. The idea is to introduce temporal information through the samples' connectivity in the embeddings. Figure~\ref{f.intro} depicts the whole pipeline.

Experiments conducted over the AV ChiME3 dataset compare the proposed approach against a CCA-based multilayer perceptron (MLP). Results show that (i) the multimodal CCA-GNNs produce more representative features than the standard unimodal version, leading to lower errors over clean audio data reconstruction; (ii) the proposed prior-frame approach for sequential-time modeling in graphs outperform the standard feature-space distance-based neighborhood connections; and (iii) CCA-GNNs deliver better results than the CCA-MLP model in the context of feature extraction for data reconstruction, requiring a considerably reduced rate of firing neurons, indicating the new model is more suitable for energy-constrained environments, such as AV hearing aid devices. To the best of our knowledge, this study is the first to demonstrate the application of GNN-CCA for energy-efficient rich real-world multi-modal data for a benchmark AV speech enhancement problem, where multiple real-world noises corrupt speech in real-world conditions. Therefore, the main contributions of this paper are threefold:

 \begin{enumerate}
	\item Demonstrate the application of a novel self-supervised energy-efficient CCA-GNN-based model that considers the fusion between different sources of information from the environment to improve sound quality.
	\item A novel GNN-CCA model with integrated positional node encoding considering a prior-frame sequence distance instead of a feature-space representation when computing the node’s nearest neighbors, introducing temporal information in the embeddings through the neighborhood’s connectivity.
	\item The proposed method is evaluated with the benchmark AV Grid and ChiME3 corpora, with $4$ different real-world noise types (cafe, street junction, public transport (BUS), pedestrian area) and compared with standard GNNs and MLP models for unsupervised AV speech processing tasks. Comparative results show that our new method demonstrates superior energy consumption and generalization performance in all experimental conditions. Other comparative models fail to reconstruct speech-in-noise with a similar number of neurons. 
 \end{enumerate}

The remainder of this paper is described as follows. Section~\ref{s.proposed} provides a brief background regarding GNN-CCA and introduces the proposed approaches. Further, Section~\ref{s.methodology} provides the necessary information regarding the dataset and the evaluation metrics, while Section~\ref{s.setup} describes the experimental setup. Finally, Sections~\ref{s.experiments} and~\ref{s.conclusions} state the results and conclusions, respectively.

\section{Related Work}
\label{sec:related}

Ngiam et al.~\cite{ngiam2011multimodal} showed that multimodal learning approaches could be beneficial to improve one modality feature representation. The approach was recently applied to a wide variety of applications, such as emotion analysis~\cite{jia2021multimodal}, scene change detection~\cite{santana2019novel} and medicine~\cite{venugopalan2021multimodal}, to cite a few. Regarding audio-visual (AV) data processing, Adeel et al.~\cite{adeel2018real} suggested an integration of Internet of Things (IoT) and 5G Cloud-Radio Access Network to create a chaotic encryption-based lightweight model for lip-reading driven hearing aids. In further work~\cite{adeel2020novel} the model was improved to transmit encrypted compressed audio-visual (AV) information and receive encrypted enhanced reconstructed speech in real-time. Recent works comprise a deep learning-based framework for speech enhancement that exploits AV cues concerning different operating conditions to estimate clean audio~\cite{adeel2020contextual}, as well as the CochleaNet~\cite{gogate2020cochleanet}, which integrates noisy audio and visuals from distinct language speakers. 

Regarding the proposed approaches for correlated feature extraction using energy-based approaches, one can refer to~\cite{hinton2002training,passos2017fine,passos2019kappa,passos2018temperature}. In the context of mutual information, Belghazi et al.~\cite{belghazi2018mine} proposed the so-called MINE, a mutual information neural estimation, while Hjelm et al.~\cite{hjelm2018learning} proposed a model for Learning deep representations by mutual information estimation and maximization. Further, Veli{\v{c}}kovi{\'c} et al.~\cite{velivckovic2018deep} proposed a graph-based solution called Deep Graph Infomax, which combines on mutual information maximization with graph neural networks.

Graph theory describes strong architectures capable of modelling complex relationships, with applications ranging from small world design~\cite{newman2000models} to oversampling~\cite{passos20202pf,passos2022handling}. Considering GNNs applications for correlated feature extraction, aside from~\cite{velivckovic2018deep}, Zhang et al.~\cite{zhang2021canonical} introduced canonical correlation analysis for self-supervised feature extraction using GNNs, providing a more efficient alternative for the task since it does not rely on negative pairs, does not require learning parameters of additional components such as an estimator. Further, the complexity of the model is considerably smaller since it requires O($F^2$) space cost against O($N$) from the Deep Graph Infomax, where $F$ denotes the feature space size and $N$ denotes the number of nodes.

\section{Graph Neural Network with Canonical Correlation Analysis}
\label{s.CCAGNN}

\begin{figure*}[!ht]
  \centerline{
    \begin{tabular}{c}
	\includegraphics[width=.9\textwidth]{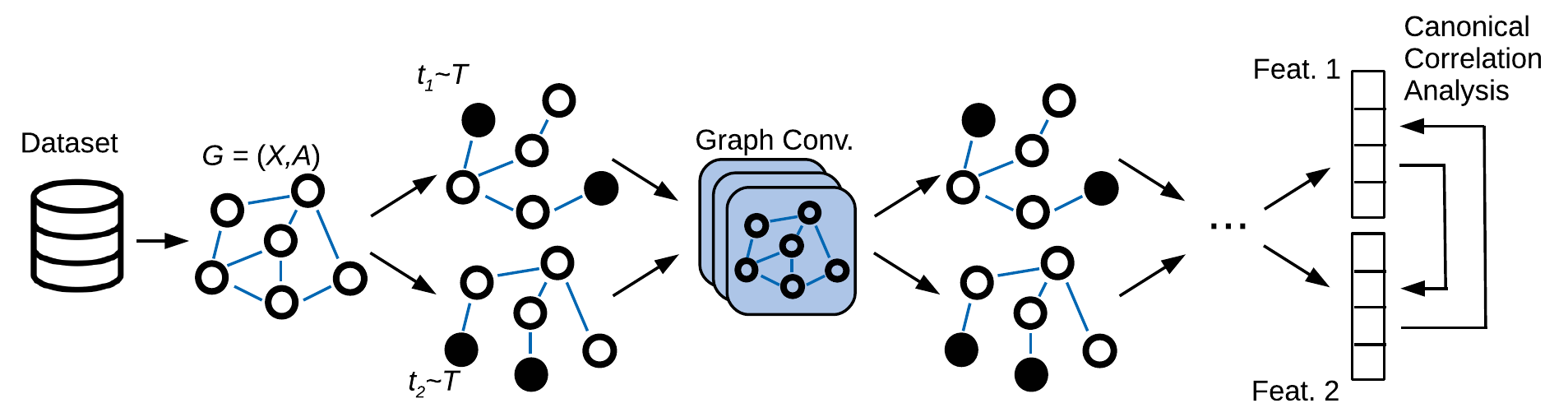} 
    \end{tabular}}
    \caption{Graph Neural Network with Canonical Correlation Analysis. The dataset is converted into a graph where each sample represents a node, and the edges denote the nodes' connection. Two augmented versions of this graph are generated and used to feed a GNN. Finally, the canonical correlation between the output of both versions is computed and used as the cost function to optimize the GNN parameters.}
  \label{f.CCAGNN}
\end{figure*}

Consider a single graph $G = (X, A)$ where $X\in \mathbb{R}^{N\times F}$ denotes the node's feature vectors and $A \in \mathbb{R}^{N \times N}$ stands for the adjacency matrix. The CCA-GNN~\cite{zhang2021canonical} is composed of three main parts: (i) a random graph generator $T$, (ii) a graph neural network encoder $f_\theta$, where $\theta$ stands for the learnable parameters, and (iii) a Canonical Correlation Analysis-based objective function. The idea is to present two augmented versions of the same graph to the network and maximize the canonical correlation between their outputs. Such an approach aims at preserving correlated components while discarding decorrelated ones, i.e., maintaining the relevant information present in both augmented versions and avoiding particular behaviors, such as anomalies and noise. Figure~\ref{f.CCAGNN} depicts the Graph Neural Network with Canonical Correlation Analysis.

Regarding the graph augmentation, CCA-GNN employs the same approach used in~\cite{zhu2020deep,thakoor2021bootstrapped}, which basically performs a random edge dropping and feature masking. Thus, each $t\sim T$ stands for a transformed version of $G$. Notice that those augmented versions are sampled at each iteration.

Concerning the encoder, the model employs a simple two-layered graph neural network, which can be easily replaced by more complex or sophisticated architectures.

Finally, the objective function aims at modeling the learning problem as a canonical correlation-based~\cite{chang2018scalable} self-supervised approach in which the two randomly augmented versions of the graph yields two normalized views of the input data, $\bm{Z}_A$ and $\bm{Z}_B$, and their correlation is maximized. The objective function is described as follows:
\begin{equation}
{\cal L}(\bm{Z}_A, \bm{Z}_B) = ||\bm{Z}_A - \bm{Z}_B||_{F}^2+\lambda\left(||\bm{Z}_A^T\bm{Z}_A-\bm{I}||_F^2+||\bm{Z}_B^T\bm{Z}_B-\bm{I}||_F^2\right), 
\label{e.canonicalCorrelation}
\end{equation}
where $I$ is the identity matrix and $\lambda$ is a non-negative trading-off hyperparameter. The first term, namely the invariance term, is responsible for the minimization of the invariance between the two views, which is essentially the same as maximizing the correlation between them. The second is the decorrelation term, which seeks a regularization that encourages distinct features to capture different semantics.

Further, the authors provided a variance-covariance perspective~\cite{tian2021understanding} of the objective function. Suppose $\bm{s}$ as an augmented version sampled from an input $\bm{x}$, and $\bm{z_s}$ is the representation of $\bm{s}$ obtained through a decoder. The invariance term can be minimized using expectation as follows:
\begin{eqnarray}
{\cal L}_{inv} &=& ||\bm{Z}_A - \bm{Z}_B||_{F}^2= \sum_{i=1}^N\sum_{k=1}^D(z_{i,j}^A-z_{i,j}^B)^2\nonumber\\
 &\cong& \mathbb{E}_{\bm{x}}\left[\sum_{k=1}^D\mathbb{V}_{\bm{s}|\bm{x}}[\bm{z_s},k]\right]*2N,
\label{e.loss_inv}
\end{eqnarray}
where $\mathbb{V}$ is the variance. In a similar fashion, the decorrelation term is written as follows:
\begin{eqnarray}
{\cal L}_{dec} &=& ||\bm{Z}_S^T\bm{Z}_S-\bm{I}||_F^2= ||\bm{Cov}_{\bm{s}}[\bm{z}]-I||_F^2\nonumber\\
 &\cong& \sum_{i\neq j}(\rho_{i,j}^{\bm{z_s}})^2,\forall \bm{Z}_S\in\{\bm{Z}_A,\bm{Z}_B\},
\label{e.loss_dec}
\end{eqnarray}
where $\bm{Cov}$ is the covariance matrix and $\rho$ is the Pearson correlation coefficient.

\section{Proposed Approach}
\label{s.proposed}

This section presents a multimodal extension for the Canonical Correlation Analysis Graph Neural Network. Further, it also introduces the idea of modeling the temporal information of sequence data as node relationships in a graph.

\subsection{Multimodal Canonical Correlation Analysis Graph Neural Network for Audio-Visual Embedding Learning}
\label{ss.CCAGNN_av}

\begin{figure*}[!htb]
  \centerline{
    \begin{tabular}{c}
	\includegraphics[width=\textwidth]{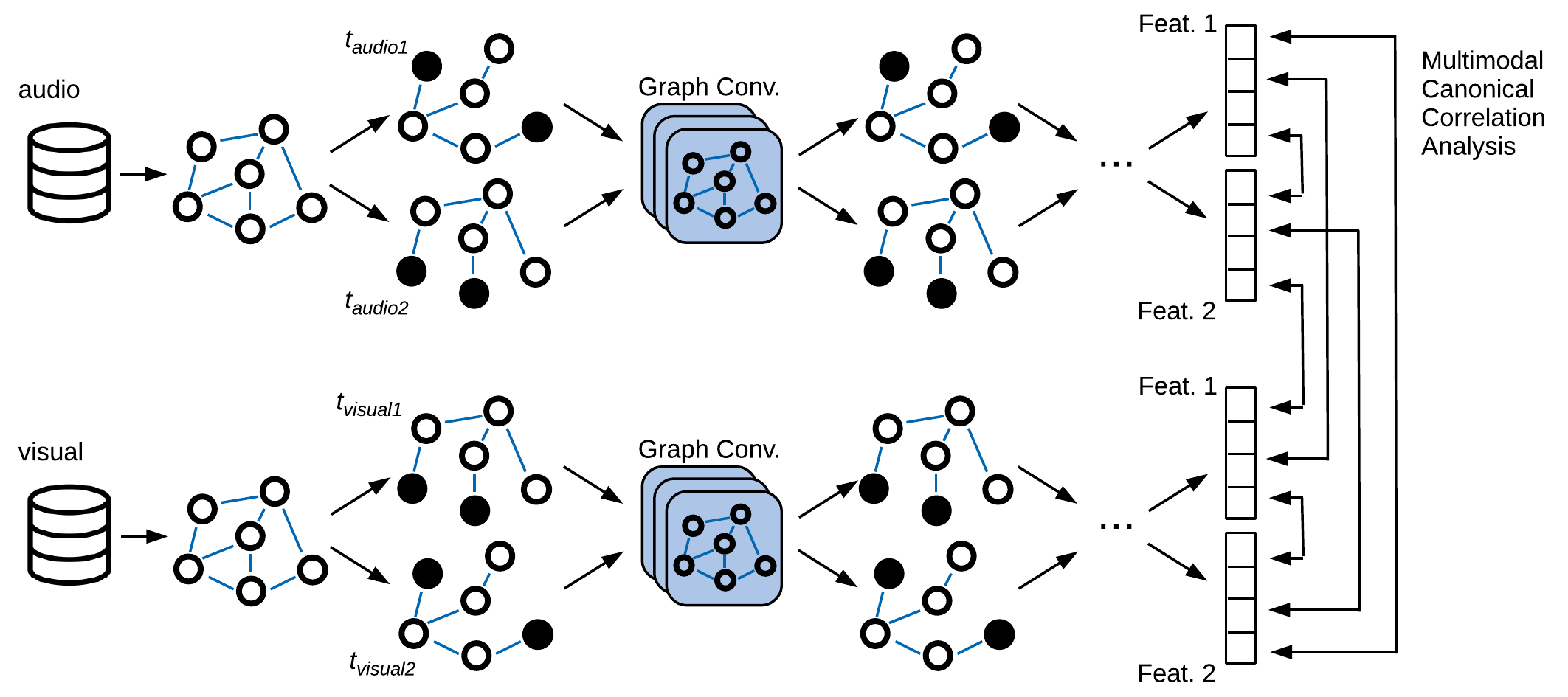} 
    \end{tabular}}
    \caption{Multimodal Canonical Correlation Analysis Graph Neural Network for Audio-Visual Embedding Learning.}
  \label{f.CCAGNN_av}
\end{figure*}

The proposed extension of the Canonical Correlation Analysis Graph Neural Network for multimodal data comprises a pair of networks, each of them fed with a modality of data, e.g., audio and visual, running in parallel. At the output layer, the canonical correlation analysis is performed considering both the intra-channel correlation, i.e., the two augmented versions of the same data modality, as well as inter-channels correlation. Figure~\ref{f.CCAGNN_av} depicts the model.

To accommodate the intra- and inter-channel computations of the canonical correlation analysis in the objective function, one can firstly consider two randomly augmented versions of normalized views for each channel, namely $\bm{Z}_1$ and $\bm{Z}_2$ for audio data, and $\bm{Z}_3$ and $\bm{Z}_4$ for visual data. The individual losses for both channels are computed using Equation~\eqref{e.canonicalCorrelation}:

\begin{equation}
\begin{array}{c}
        {\cal L}_{\text{Audio}} =  {\cal L}(\bm{Z}_1, \bm{Z}_2),\\ 
        {\cal L}_{\text{Visual}} =  {\cal L}(\bm{Z}_3, \bm{Z}_4).
\end{array}
\label{e.AudioVisual_alternative}
\end{equation}
 
Further, all the possible combinations of audio and visual data are computed, namely \textit{Audio1Visual1} ({$Z_1,Z_3$}), \textit{Audio1Visual2} ({$Z_1,Z_4$}), \textit{Audio2Visual1} ({$Z_2,Z_3$}), and \textit{Audio2Visual2} ({$Z_2,Z_4$}), as follows:

\begin{equation}
\begin{array}{c}
        {\cal L}_{\text{Audio1Visual1}} =  {\cal L}(\bm{Z}_1, \bm{Z}_3),\\ 
        {\cal L}_{\text{Audio1Visual2}} =  {\cal L}(\bm{Z}_1, \bm{Z}_4),\\ 
		{\cal L}_{\text{Audio2Visual1}} =  {\cal L}(\bm{Z}_2, \bm{Z}_3),\\ 
		{\cal L}_{\text{Audio2Visual2}} =  {\cal L}(\bm{Z}_2, \bm{Z}_4).
\end{array}
\label{e.combinedLosses}
\end{equation}

Finally, the objective function of the multimodal Canonical Correlation Analysis Graph Neural Network is given by:
\begin{eqnarray}
{\cal L} &=&  \alpha{\cal L}_{\text{Audio}} + \beta{\cal L}_{\text{Visual}}\nonumber\\  
&& + \gamma({\cal L}_{\text{Audio1Visual1}}+{\cal L}_{\text{Audio1Visual2}}\nonumber\\  
&&   +{\cal L}_{\text{Audio2Visual1}}+{\cal L}_{\text{Audio2Visual2}}),
\label{e.finalLoss}
\end{eqnarray}
where $\alpha$, $\beta$, and $\gamma$ are constants that control the influence of audio, video, and the combined canonical correlation, respectively.

\subsection{Modelling Temporal Information as Graph Nodes Relationships}
\label{ss.temporalRelationship}

\begin{figure*}[!htb]
  \centerline{
    \begin{tabular}{cc}
	\includegraphics[width=.44\textwidth]{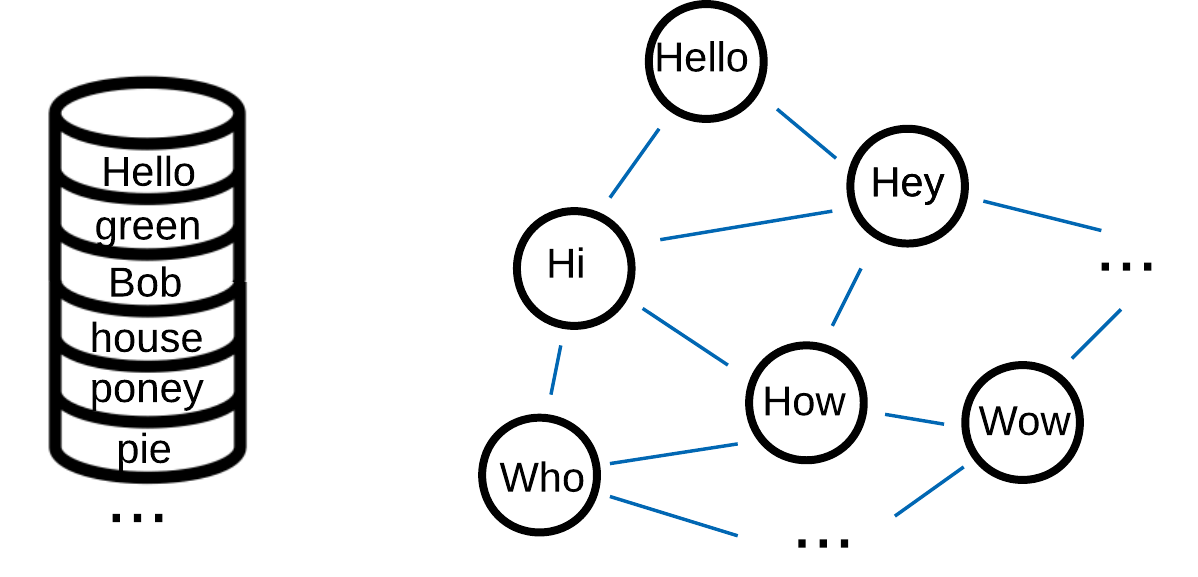} &
	\includegraphics[width=.44\textwidth]{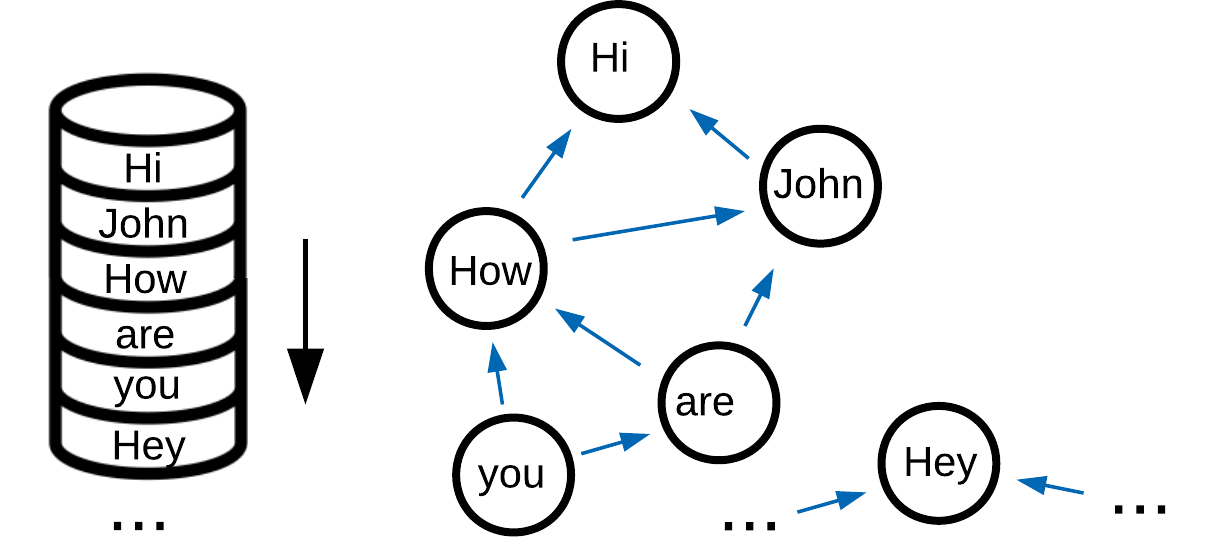} \\
	(a) & (b)
    \end{tabular}}

    \caption{Node neighborhood modeling considering (a) the standard feature-space-based $k$-nearest neighbors approach and (b) the proposed $k$ prior frames with $k=2$.}
  \label{f.sequentialGraph}
\end{figure*}

The usual approach for modeling a dataset into a graph structure consists of representing its samples as the graph's nodes, whose edges connect the adjacent instances inserted into a $D$-dimensional feature space. A common approach is to connect each node to its $k$ nearest neighbors only, where $k$ is a hyperparameter, presenting two main advantages: (i) it reduces the computational burden since it considers only $k$ operations per node instead of $N$, and (ii) it enhances the influence of the neighborhood of the node, avoiding the effect of uncorrelated samples to the local process. Figure~\ref{f.sequentialGraph}(a) depicts the idea. 

This paper proposes a novel approach for modeling the nodes' connectivity considering temporal information propagation instead of the distance in the feature space. The strategy conducts the positional encoding of the instances by connecting each vertex to its $k$ previous nodes, e.g., frames in a video sequence. Figure~\ref{f.sequentialGraph}(b) illustrates the process.

Notice that the edges between each pair of connected nodes are weighted accordingly to their distances in this temporal representation, i.e., the first prior frame of a node is more strongly connected to it than the second prior, and so on consecutively. The edge weight $w_{ij}$ connecting a node $i$ to a previous $j$ is computed as follows:
\begin{equation}
w_{ij}=k+1 - d_{ij},
\label{e.weight_distribution}
\end{equation}
where $d_{ij}$ is the distance from $i$ to $j$ in prior frames steps, i.e., $d_{ij}=1$ means $j$ is the first prior frame of  $i$, while $d_{ij}=2$ means $j$ is two prior frames away from $i$, and so on. Notice each node is also connected to itself through a self-reference edge $w_{ii} = k+1$ since $d_{ii}=0$. Moreover, the experiments also consider weighting self-reference connections $w_{ii} = 1$, increasing the influence of neighborhood in the GNN decision process. After defining the distance between each pair of connected samples, such values are stored in a distance matrix employed to compute nodes' normalized positional encoding, replacing the adjacency matrix in the factorization of the graph Laplacian.

In a nutshell, the multimodal approach and the positional encoding complement each other with specific helpful information. The multimodal architecture aims to maximize the canonical correlation between audio and visual channels, whose objective is amplifying the flow of correlated information associated with the speech and suppressing uncorrelated information related to noise. In other words, it focuses on the synchronicity of the sounds and the mouth movements. On the other hand, positional encoding aims to introduce a temporal dependence in this information since meaning in speech is associated with prior sounds, words, and sentences.

\section{Methodology}
\label{s.methodology}

This section describes the dataset considered for the task of audio/visual correlated embedding learning for the task of clean sound reconstruction and the process employed for feature extraction. Further, it also exposes the evaluation metrics considered in the experiments.

\subsection{AV ChiME3 Dataset}
\label{ss.dataset}

This paper employs a dataset composed of pairs of image and noisy audio signals for input and clean audio signals for output, aiming to provide an efficient tool capable of enhancing and cleaning the relevant audio signal considering environmental information fusion. The dataset comprises a combination of clean videos from the Grid~\cite{cooke2006audio} dataset with noises (pedestrian area, public transport, street junction, cafe) with signal to noise ratios (SNR) ranging from -12 to 12dB extracted from ChiME3~\cite{barker2015third}, composing the AV ChiME3~\cite{adeel2019lip} dataset. The preprocessing comprises sentence alignment, which is conducted to prevent the model from learning redundant or insignificant information and removing silent takes from data, as well as incorporating prior multiple visual frames used to include temporal data, thus improving the mapping between audio and visual characteristics. The dataset comprises $5$ speakers (one black male, two white males, and two white females) selected from Grid corpus reciting $989$ sentences each. 

\subsubsection{Audio feature extraction}
\label{sss.audioDataset}
The audio features are extracted using log-FB vectors, which are computed by sampling the input audio signal at $22,050$kHz and segmented into $M$ $16$ms frames with $800$ samples per frame. Notice that each instance is sampled using a hamming window function~\cite{bojkovic2017hamming} with $62.5\%$ of the frame size ($500$ samples), therefore performing some frame overlapping during the procedure steps. Further, the Fourier transform is computed to produce a 2048-bin power spectrum. Finally, a logarithmic compression is applied to obtain a  $22$-dimensional log-FB signal.

\subsubsection{Visual feature extraction}
\label{sss.visualDataset}
The visual features were extracted from the Grid Corpus dataset through a simple encoder-decoder setup approach. After extracting a sequence of individual frames, the lip-regions are identified using Viola-Jones~\cite{viola2001rapid} and tracked across a sequence of frames using a method proposed in~\cite{ross2008incremental}. The sentences are manually inspected using a random approach to ensure good lip tracking and delete sentences with misclassified lip regions~\cite{abel2016data}. Finally, the encoder-decoder approach is employed to produce vectors of pixel intensities, whose first $50$ components are vectorized in a zigzag order and then interpolated to match the equivalent audio sequence. 

\begin{figure}[!htb]
  \centerline{
    \begin{tabular}{c}
	\includegraphics[width=.6\textwidth]{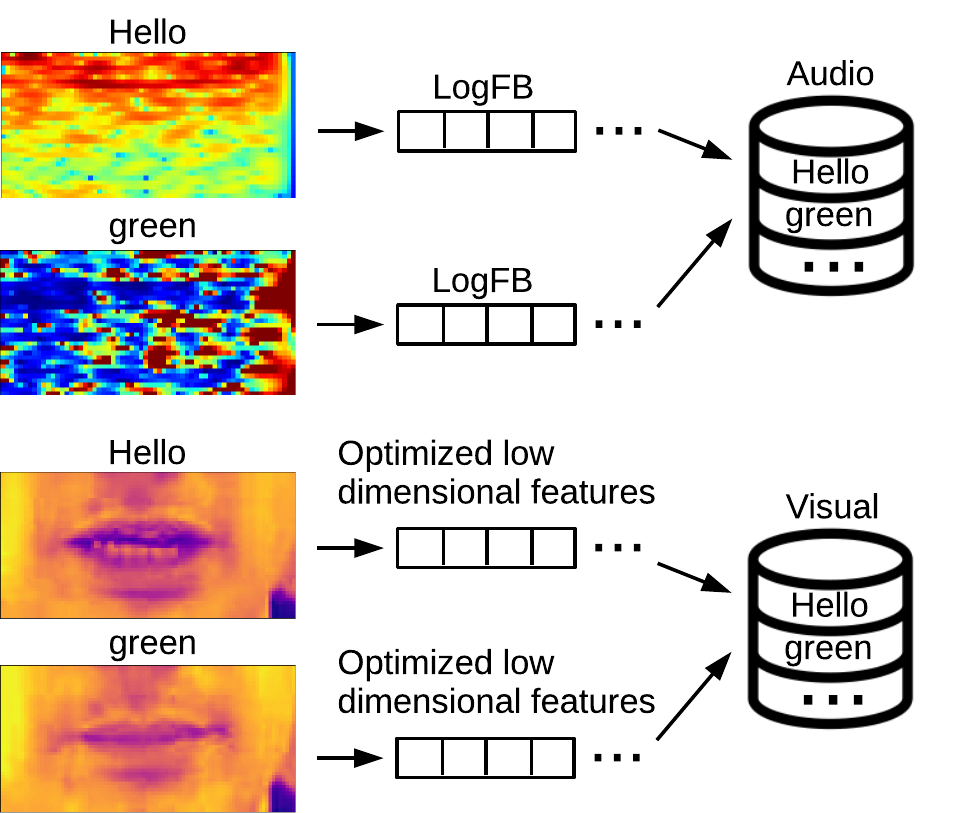} 
    \end{tabular}}
    \caption{Simplified feature extraction process schema.}
  \label{f.dataset}
\end{figure}

Finally, the dataset employed in this paper is composed of three subsets, i.e., clean audio, noisy audio, and visual features. Both clean and noisy audio subsets comprise $22$ features each, while the visual features are represented by a $50$-dimensional vector. The subsets contain $750$ out of $989$ sequences with $48$ frames each, summing up to a total of $36,000$ synchronized samples per subset. Figure~\ref{f.dataset} illustrates a simplified schema of the feature extraction process.

\subsection{Evaluation Metrics}
\label{ss.metrics}

This section provides a brief description of the metrics employed to evaluate the experiments, i.e., Perceptual Evaluation of Speech Quality (PESQ), Mean Square Error (MSE), Area under the Curve, and energy efficiency.

\subsubsection{PESQ}
\label{sss.PESQ}

Perceptual Evaluation of Speech Quality (PESQ) comprises a set of mechanisms for automated speech quality evaluation~\cite{recommendation2001perceptual}. It was developed for objective voice quality testing by telecom operators, equipment vendors, and phone manufacturers.  

Its testing topology depends on the available information, and the method is divided into two classes:

\begin{itemize}
	\item Full Reference (FR): uses the original signal as a reference for comparison. This approach compares each reference sample to the corresponding noisy signal, thus delivering more accurate results.
	\item No Reference (NR): uses only the noisy signal for quality estimation and has no information regarding the reference signal. 
\end{itemize}

In a nutshell, PESQ is a full-reference algorithm that analyzes the speech signal sample-by-sample after a temporal alignment of corresponding excerpts of reference and test signal. This paper employs the FR version.

\subsubsection{MSE}
\label{sss.MSE}

The Mean Squared Error measures the average of the squares of the errors, i.e., the average squared difference between the estimated values and the actual value. The MSE formulation is described as follows:

\begin{equation}
MSE = \frac{1}{n}\sum_{i=1}^n(y_i-\hat{y}_i)^2,
\label{e.mse}
\end{equation}
where $n$ stands for the number of data points, $y$ is the expected value, and $\hat{y}$ is the predicted value.

\subsubsection{Area under the Curve}
\label{sss.AuC}

This work employs a discrete version of the area under the curve, which denotes the sum of the outputs of a given function $f(x)$ such that $\{x\in\mathbb{R}|a\leq x \leq b\}$, where $a$ and $b$ denote the lower and upper bounds, respectively. One can define the area under the curve as follows:

\begin{equation}
Area = \sum_{x=a}^bf(x).
\label{e.auc}
\end{equation}

\subsubsection{Energy Efficiency}
\label{sss.energyEfficiency}

This paper employs the term ``energy efficiency'' to describe the rate of active neurons during execution. This approach considers the idea that neurons that do not ``fire'' during the execution of the model are not performing any costly operation, thus saving energy. In a nutshell, the algorithm computes the sum of the activations whose output is greater than zero and divides it by the total number of activations. 

\section{Experimental Setup}
\label{s.setup}

The experiments provided in the next section were conducted considering a graph neural network and a multilayer perceptron network as the backbone. Both networks share a similar architecture for comparison purposes, i.e., two hidden layers with $512$ neurons in each layer, using the Adam optimizer with a learning rate of $0.001$. The models are trained with the objective of maximizing the canonical correlation for coherent features extraction during $5,000$ epochs considering a trading-off parameter $\lambda=0.0001$ in Equation~\eqref{e.canonicalCorrelation}, while Equation~\eqref{e.finalLoss} is set with the values $\alpha=0.5$, $\beta=0.25$, and $\gamma=0.0625$. Notice that these values were empirically chosen using a grid search in the range $[0, 1]$ with step $0.0625 (1/16)$. These values make sense since they give more importance to the receptive input than to the context provided by the visual information and a combination of both, and the most relevant information to reconstruct the clean audio in this context is the noisy audio. The data augmentation step considers both an edge dropping rate and a feature masking rate of $0.5$. Moreover, the graphs are generated considering three distinct neighborhood scenarios, i.e., with $3$, $10$, and $30$ neighbors. Notice this neighborhood is defined in two manners, i.e., the standard feature-space-based approach and the proposed temporal-information-based relationship, as described in Section~\ref{ss.temporalRelationship}.

After training, the networks' outputs are used to feed a dense layer, which is responsible for reconstructing the clean signal given the features extracted from noisy audio for the single modality and noisy audio and clean video for the proposed multimodal extension. The dense layer is optimized during $600$ epochs using the Adam optimizer with a learning rate of $0.005$ and a weight decay of $0.0004$ using the minimization of the mean squared error as the objective function.

The dataset was divided into $15$ folds to provide an in-depth statistical analysis. Each fold comprises $50$ sequences of $48$ frames each, summing up to a total of 2,400 samples per fold. As stated in Section~\ref{ss.dataset}, the dataset is formed by three subsets: (i) clean audio, (ii) noisy audio, and (iii) clean visual. The noisy audio is used to train the standard unimodal approach, while the proposed multimodal approach employs both the noisy audio and clean visual in the training process. The clean audio is used as a reconstruction target for both cases. Finally, each fold is split into train, validation, and test sets, following the proportions of $60\%$, $20\%$, and $20\%$, respectively. For statistical evaluation, the Wilcoxon signed-rank test~\cite{Wilcoxon:45} with $5\%$ of significance was considered.

\section{Experiments}
\label{s.experiments}

This section presents the experimental results considering three tasks: (i) feature extraction driven by canonical correlation analysis maximization, (ii) clean audio data reconstruction based only on noisy audio or noisy audio and clean visual data fusion, and (iii) energy efficiency analysis in terms of neuron activation rate.

\subsection{Feature Extraction Analysis}
\label{ss.featureExtraction}

The experiments presented in this section compare the performance of graph neural networks with canonical correlation analysis for the task of self-supervised relevant feature extraction in three distinct neighborhood scenarios, i.e., $3$, $10$, and $30$ neighbors, namely GNN 3, GNN 10, and GNN 30, respectively. Moreover, two distinct neighborhood strategies are compared: (i) the standard (Std.) feature-space distance and (ii) the proposed prior frames-based connections for time-sequence (Seq.). Finally, a multilayer perceptron network with similar architecture (same number of layers, neurons per layer, optimizer, and learning rate) was trained using an identical self-supervised approach with the canonical correlation analysis maximization as the target function, denoted the baseline.

\begin{figure}[!htb]
  \centerline{
    \begin{tabular}{cc}
	\includegraphics[width=.44\textwidth]{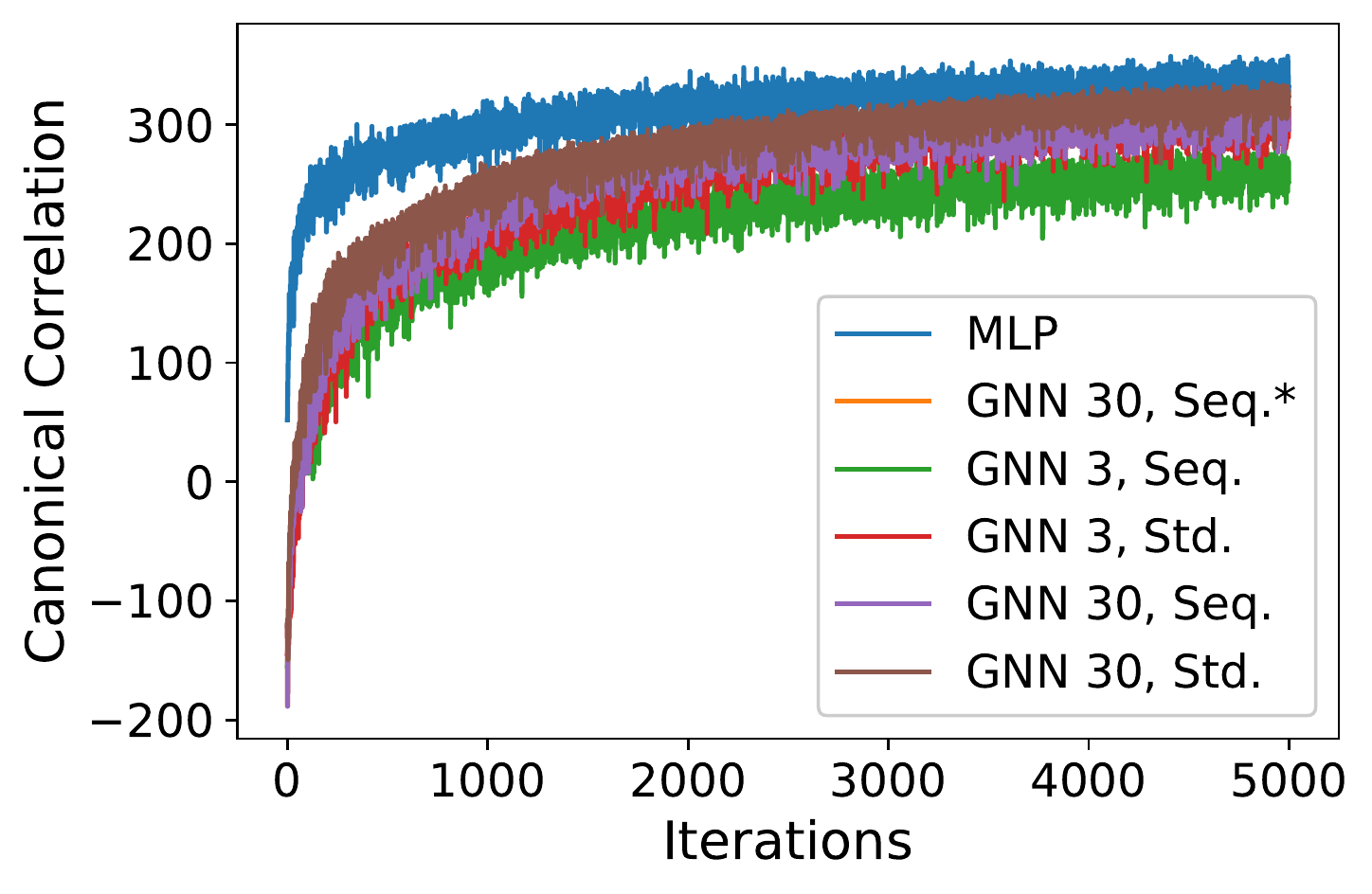} &
	\includegraphics[width=.44\textwidth]{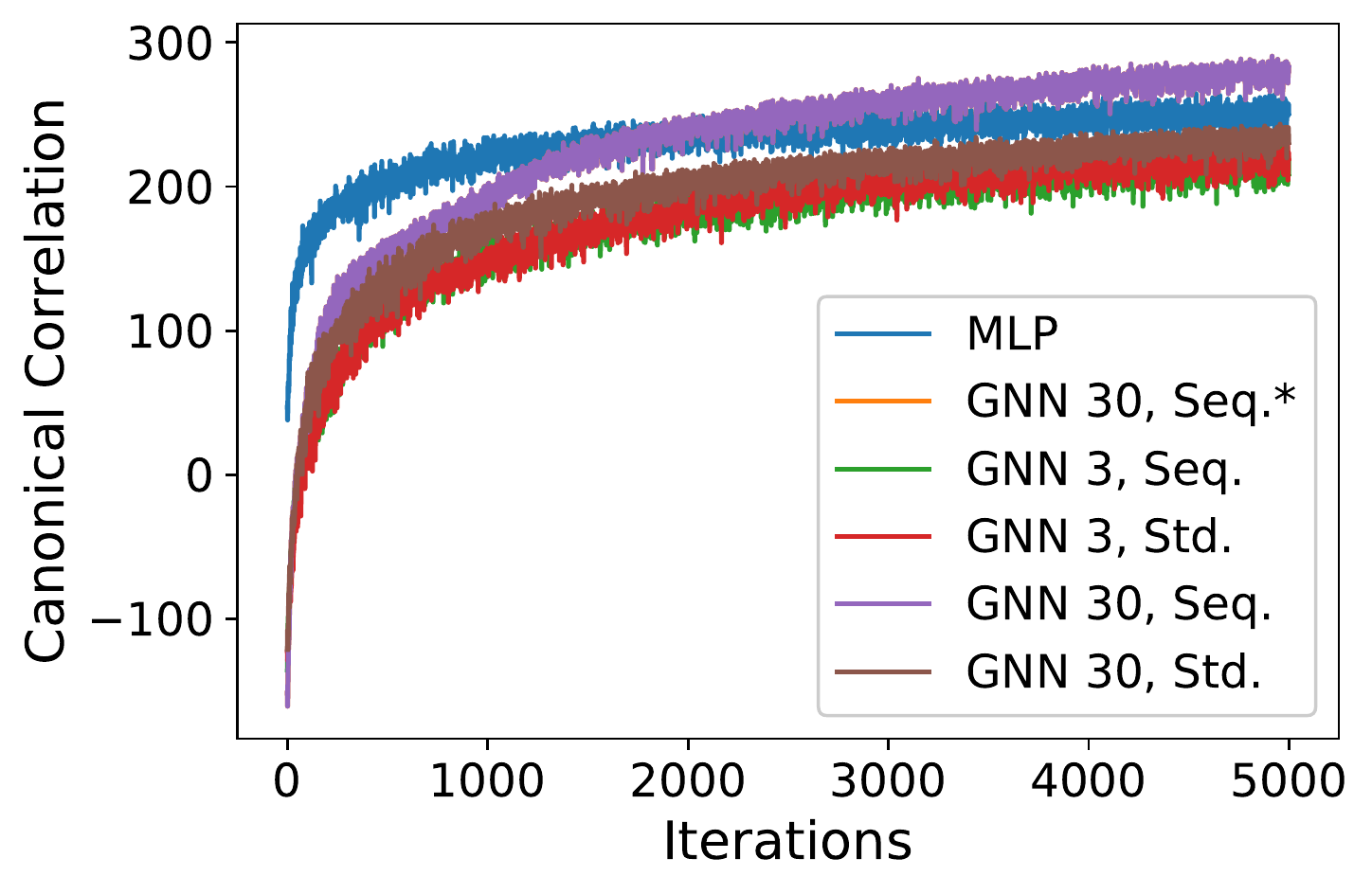} \\
	(a) & (b)
    \end{tabular}}
    \caption{Feature extraction for (a) noisy audio only and (b) noisy audio and clean visual.}
  \label{f.feature_extraction_audioOnly}
\end{figure}

Figure~\ref{f.feature_extraction_audioOnly}(a) presents the convergence of the unimodal models considering the task of feature extraction through CCA maximization over noisy audio data. In this context, one can observe CCA-MLP obtained the highest results considering CCA maximization itself, even though such results do not necessarily imply better data representation for the specific task of clean audio reconstruction since the prior frames-based CCA-GNN (Seq. and Seq.* for $w_{ii} = k+1$ and $w_{ii} = 1$, respectively) obtained better reconstruction rates, as presented in Table~\ref{t.reconstruction_audioOnly}. Notice that CCA-GNN Seq.* obtained similar results to CCA-GNN Seq. and thus is overlapped in the image.

Regarding the multimodal extension conducted over noisy audio and visual features, one can observe in Figure~\ref{f.feature_extraction_audioOnly}(b) that the proposed CCA-GNN Seq. obtained the best results overall considering a neighborhood (number of prior frames) of $30$. Such a behavior is expected since a more significant number of prior frames reinforces the coherent information shared between the two channels during a longer period of time, providing more robust and connected features. Such robustness is in fact propagated to the reconstruction task since CCA-GNN Seq. with $k=30$ also provided the best results available in Table~\ref{t.reconstruction_audioOnly}.

\subsection{Clean Signal Reconstruction}
\label{ss.reconstruction}

\begin{figure}[!htb]
  \centerline{
    \begin{tabular}{c}
	\includegraphics[width=.60\textwidth]{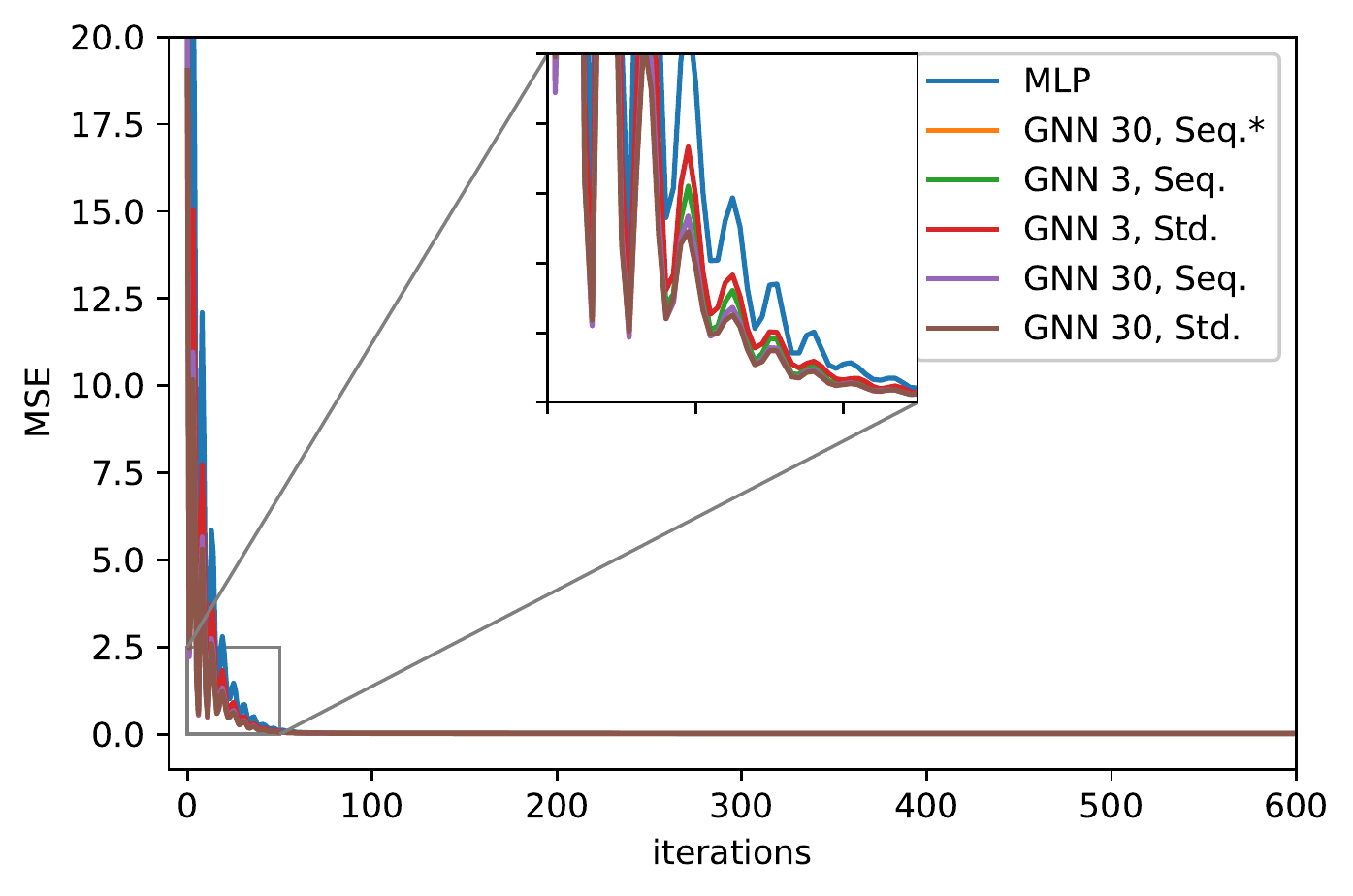} 
    \end{tabular}}
    \caption{Clean audio reconstruction error convergence considering the unimodal architecture based on the noisy audio.}
  \label{f.reconstruction_audioOnly}
\end{figure}

Table~\ref{t.reconstruction_audioOnly} presents the clean audio data reconstruction error considering the unimodal approach given the noisy data signal as input. Notice Standard denotes the common approach using the feature space to represent the nodes adjacency, while Sequential and Sequential* stand for the proposed prior frame-based approach for the positional encoding using $w_{ii} = k+1$ and $w_{ii} = 1$, respectively. From these results, one can observe that (i) a more significant number of nodes' neighbors lead to better reconstruction errors; (ii) the proposed sequential approach with prior frame connections outperformed the standard feature space distance-based neighborhood modeling and the MLP, and (iii) Sequential and Sequential* did not present significative differences, showing that the influence of the node self-connection is not relevant for the task. Notice that the better results according to the Wilcoxon signed-rank test are presented in bold. In this context, none of the other approaches obtained statistically similar results to the proposed sequential CCA-GNN with $k=30$. Further, Figure~\ref{f.reconstruction_audioOnly} shows that all techniques present similar convergence considering the reconstruction task, even though the MLP presents a slightly slower convergence, as depicted in the zoomed frame.

\begin{table}[!htb]
	\caption{Average Mean Squared Error and standard deviation over unimodal CCA-MLP and CCA-GNN considering clean audio reconstruction given noisy audio input.}
	\begin{center}
		\resizebox{0.7\textwidth}{!}{
			\begin{tabular}{cccccc}
				\toprule
				Model & Neighbors & Standard & Sequential & Sequential* \\
				\midrule
				\textbf{MLP} & - &  $0.0206\pm0.0012$ & -& - \\
				\midrule
				\multirow{3}{*}{\textbf{GNN}} & 3 & $0.0238\pm0.0009$ & $0.0220\pm0.0025$& $0.0220\pm0.0025$ \\ 
				& 10 &  $0.0235\pm0.0008$ & $0.0218\pm0.0025$& $0.0218\pm0.0025$ \\
				& 30 &  $0.0234\pm0.0008$ & \bm{$0.0187\pm0.0026$}& \bm{$0.0186\pm0.0025$}\\ 
				\bottomrule
		\end{tabular}}
		\label{t.reconstruction_audioOnly}
	\end{center}
\end{table}

\begin{figure}[!htb]
  \centerline{
    \begin{tabular}{cc}
	\includegraphics[width=.44\textwidth]{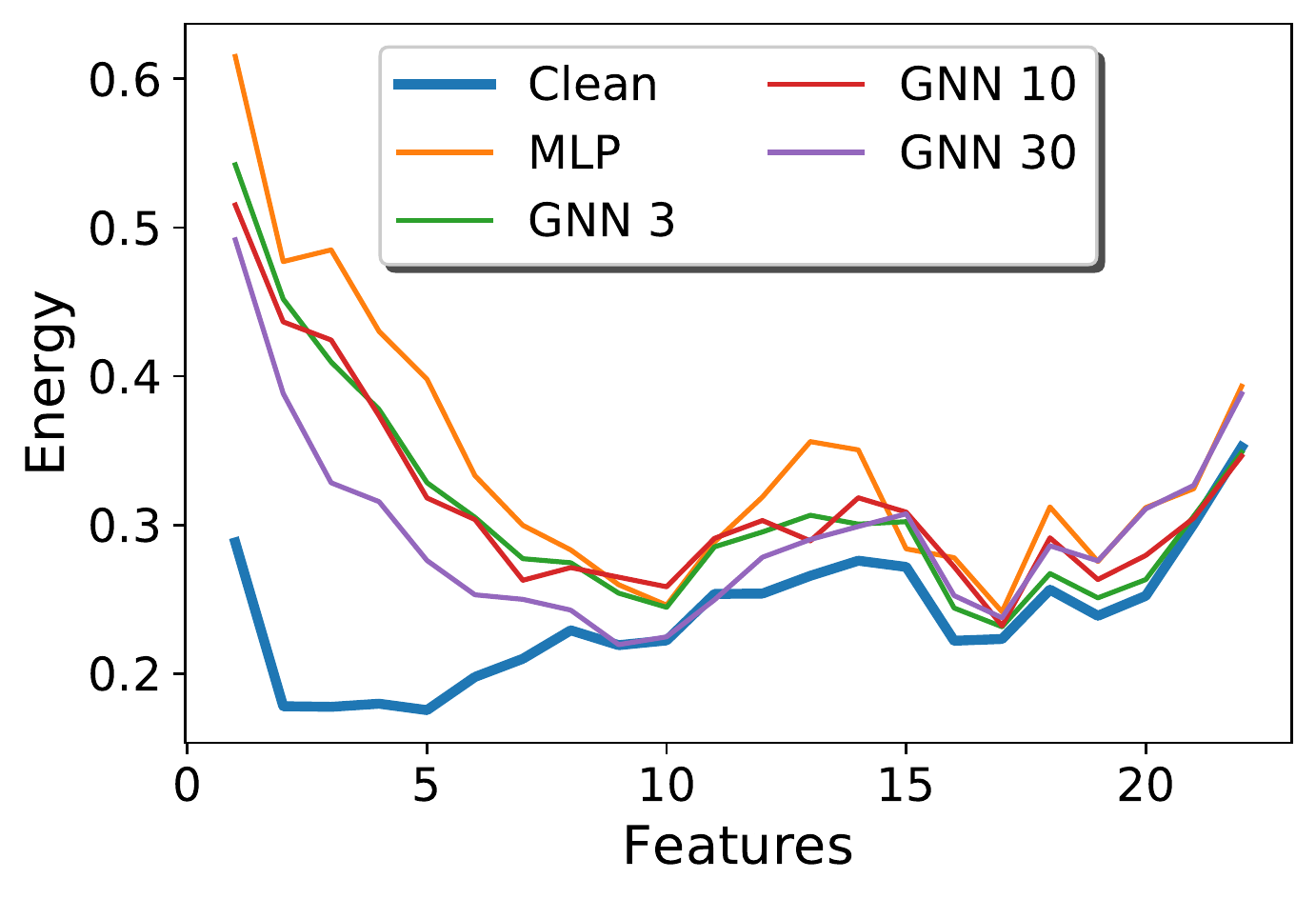} &
	\includegraphics[width=.44\textwidth]{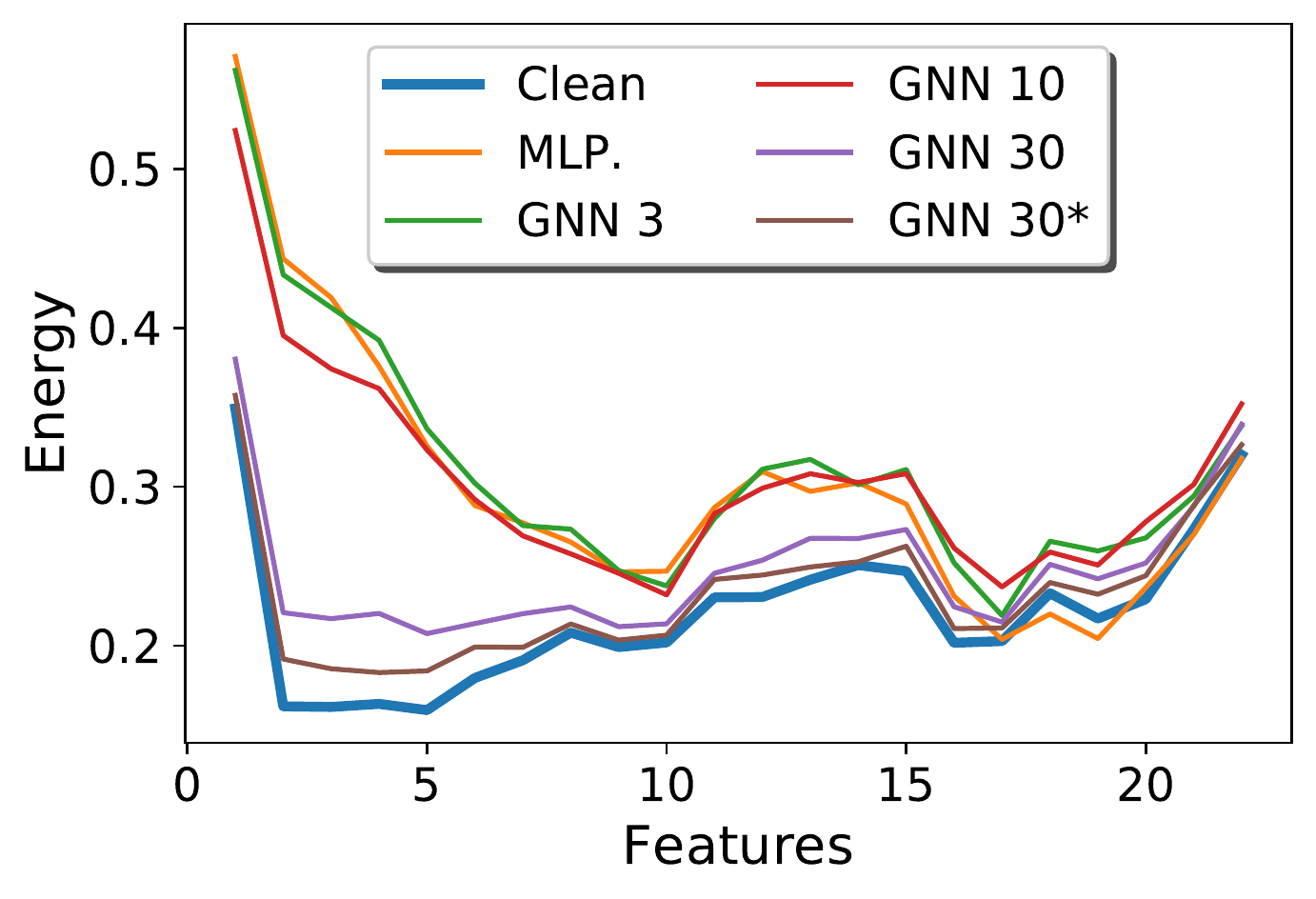} \\
	(a) & (b)
    \end{tabular}}
    \caption{Clean audio signal reconstruction regarding the unimodal architecture trained over noisy audio data considering the (a) standard approach and (b) the proposed prior frame-based positional encoding for time sequence modeling.}
  \label{f.signal_reconstruction_audio_only}
\end{figure}

Figure~\ref{f.signal_reconstruction_audio_only} depicts a randomly selected clean audio sample reconstruction regarding the unimodal architecture trained over noisy audio data considering the standard approach and the proposed prior frame-based positional encoding for time sequence modeling. Regarding the standard approach, one can notice in Figure~\ref{f.signal_reconstruction_audio_only}(a) that none of the models performed significantly well, especially for the first $8$ features. On the other hand, the proposed approach for frame-based positional encoding obtained much better results in this context considering the sequential CCA-GNN with $k=30$, as illustrated in Figure~\ref{f.signal_reconstruction_audio_only}(b). 

\begin{figure}[!htb]
  \centerline{
    \begin{tabular}{c}
	\includegraphics[width=.55\textwidth]{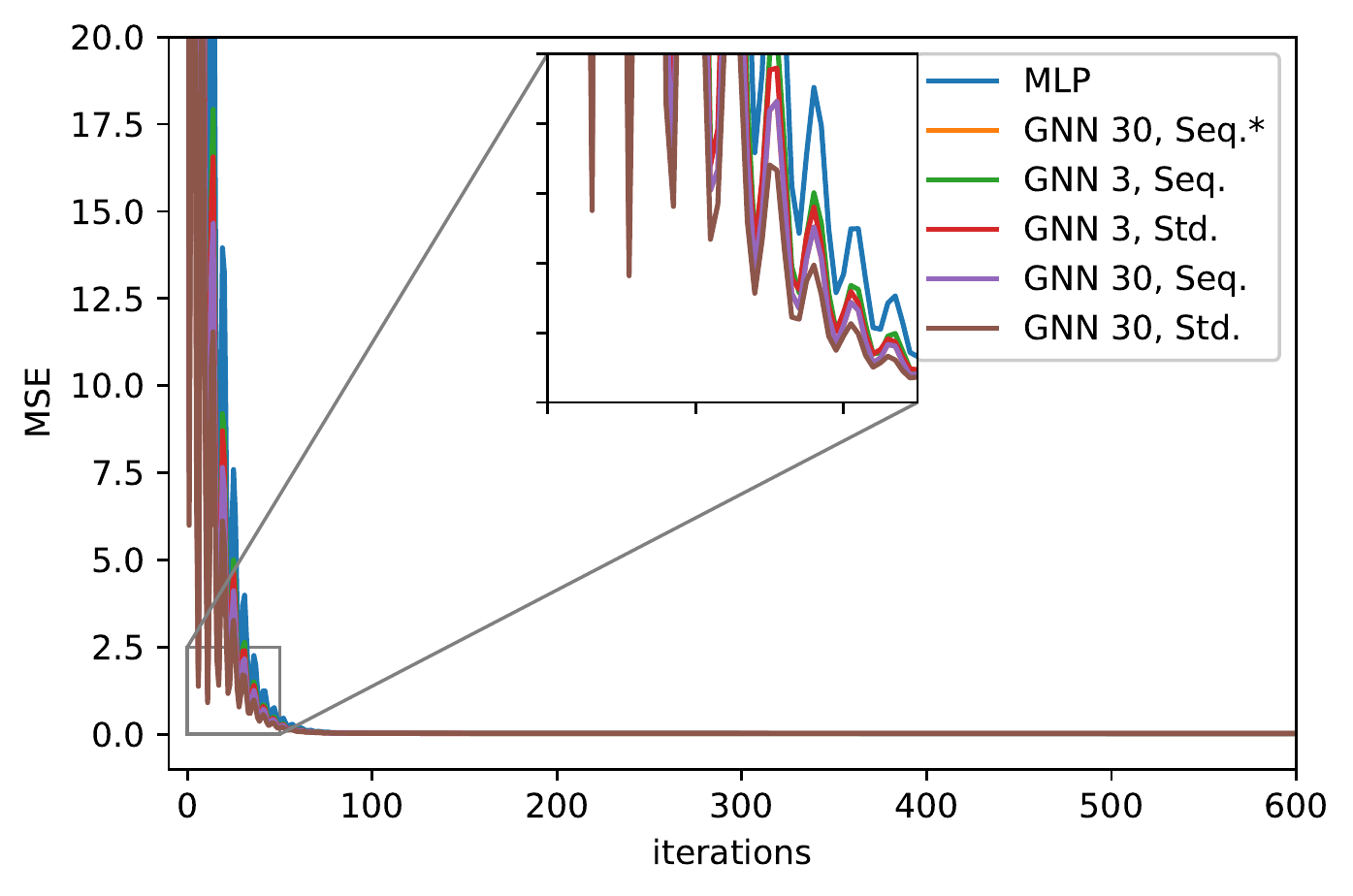} 
    \end{tabular}}
    \caption{Clean audio reconstruction error convergence considering the proposed multimodal architecture based on the noisy audio and the clean visual data.}
  \label{f.reconstruction_audioVisual}
\end{figure}

Table~\ref{t.reconstruction_audioVisual} present very similar results in the context of the multimodal architecture considering the noisy data and clean video signal as input, in which the proposed prior frame-based neighborhood outperformed the standard feature distance node's connection and the CCA-GNN with $k=30$ obtaining the better results overall considering the Wilcoxon signed-rank test. This table also shows that the multimodal approach is, in fact, capable of providing more representative features for clean audio reconstruction since all methods outperformed the respective unimodal versions. Figure~\ref{f.reconstruction_audioVisual}, which depicts the reconstruction error convergence considering the multimodal architectures, also presents very similar results to Figure~\ref{f.reconstruction_audioOnly}, in which all techniques perform similarly, with CCA-MLP showing a slower convergence in the first $50$ iterations.

\begin{table}[!htb]
	\caption{Average Mean Squared Error and standard deviation over unimodal CCA-MLP and CCA-GNN considering clean audio reconstruction given noisy audio and visual inputs.}
	\begin{center}
		\resizebox{0.7\textwidth}{!}{
			\begin{tabular}{cccccc}
				\toprule
				Model & Neighbors & Standard & Sequential & Sequential* \\
				\midrule
				\textbf{MLP} & - & $0.0189\pm0.0009$ & -& - \\
				\midrule
				\multirow{3}{*}{\textbf{GNN}} & 3 & $0.0238\pm0.0009$ & $0.0220\pm0.0025$& $0.0220\pm0.0024$ \\ 
				& 10 &  $0.0233\pm0.0009$ & $0.0216\pm0.0025$& $0.0216\pm0.0026$ \\
				& 30 &  $0.0225\pm0.0009$ & \bm{$0.0179\pm0.0025$} & \bm{$0.0180\pm0.0026$}\\ 
				\bottomrule
		\end{tabular}}
		\label{t.reconstruction_audioVisual}
	\end{center}
\end{table}


\begin{figure}[!htb]
  \centerline{
    \begin{tabular}{cc}
	\includegraphics[width=.44\textwidth]{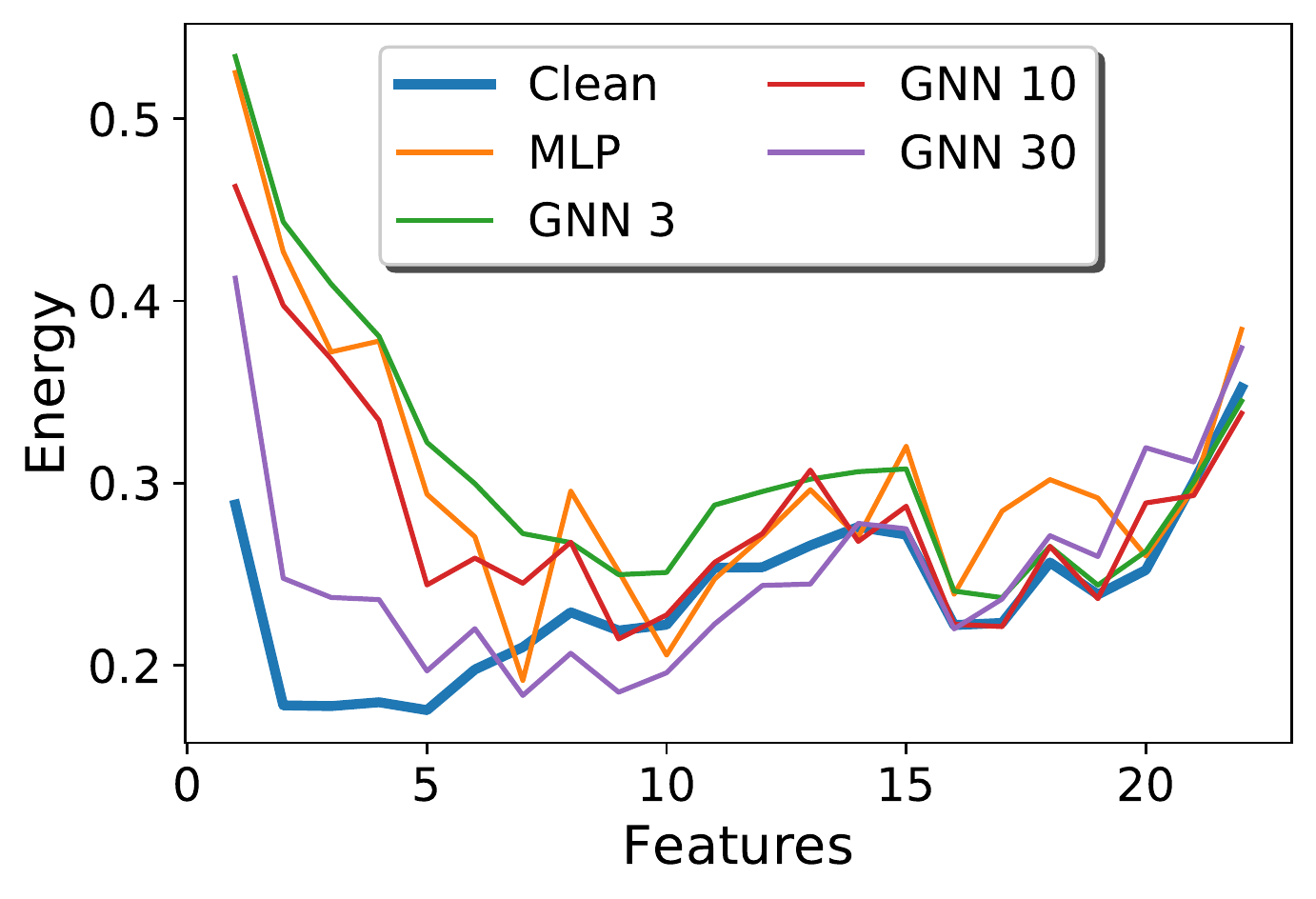} &
	\includegraphics[width=.44\textwidth]{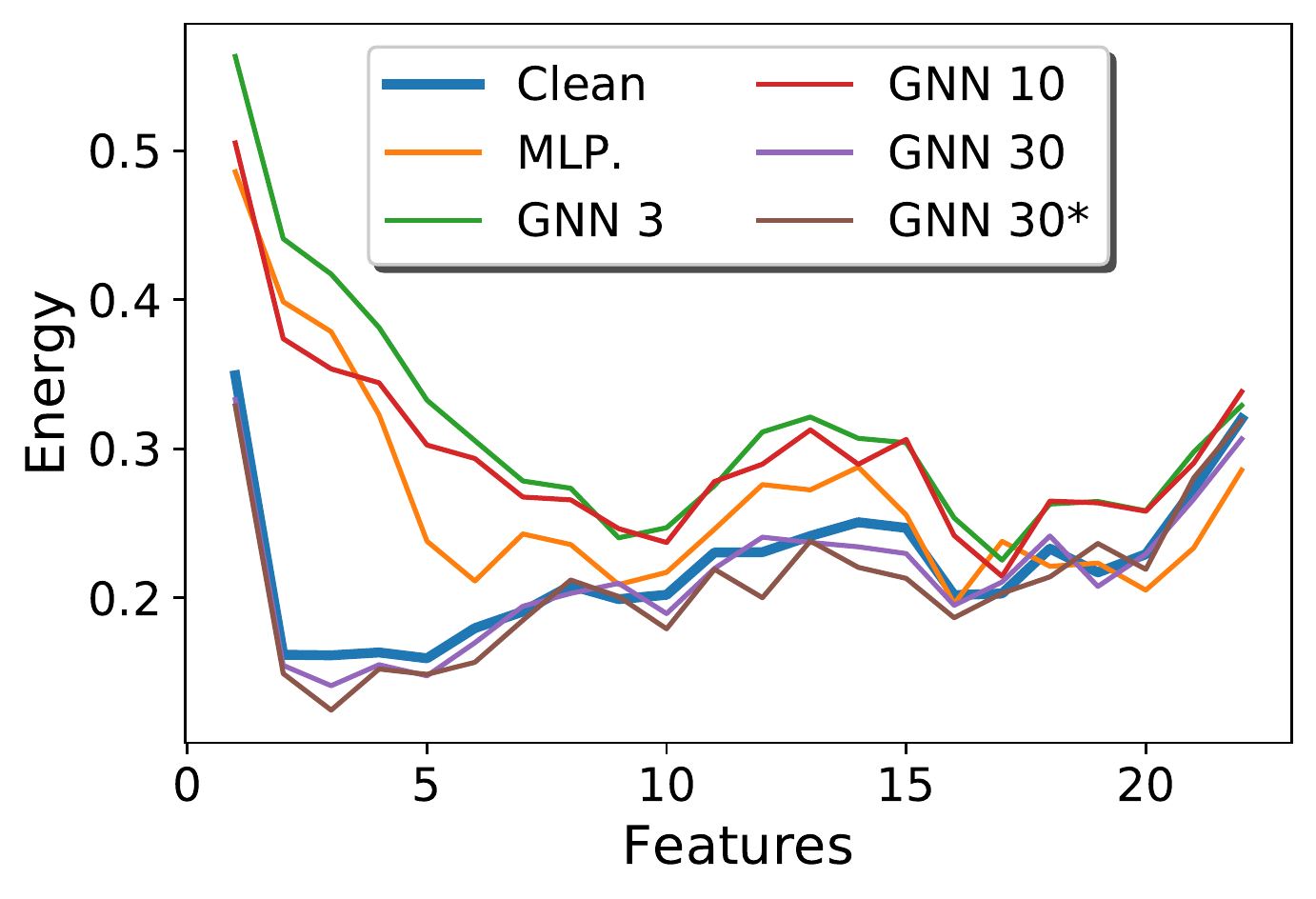} \\
	(a) & (b)
    \end{tabular}}
    \caption{Clean audio signal reconstruction regarding the multimodal architecture trained over noisy audio and clean visual data considering the (a) standard approach and (b) the proposed prior frame-based positional encoding for time sequence modeling.}
  \label{f.signal_reconstruction_audio_visual}
\end{figure}

Similar to Figure~\ref{f.signal_reconstruction_audio_only}, Figure~\ref{f.signal_reconstruction_audio_visual} presents a randomly selected clean audio sample reconstruction regarding the multimodal architecture trained over noisy audio and clean visual data. Again, Figure~\ref{f.signal_reconstruction_audio_visual}(a) shows the standard approach did not performed significantly well for the first five or eight features, while the proposed approach with $k=30$ provided much more accurate results, as depicted in Figure~\ref{f.signal_reconstruction_audio_visual}(b).

\textcolor{white}{}

\subsection{Neuronal Activation Analysis}
\label{ss.neuronActivation}

\begin{figure}[!htb]
  \centerline{
    \begin{tabular}{c}
	\includegraphics[width=.55\textwidth]{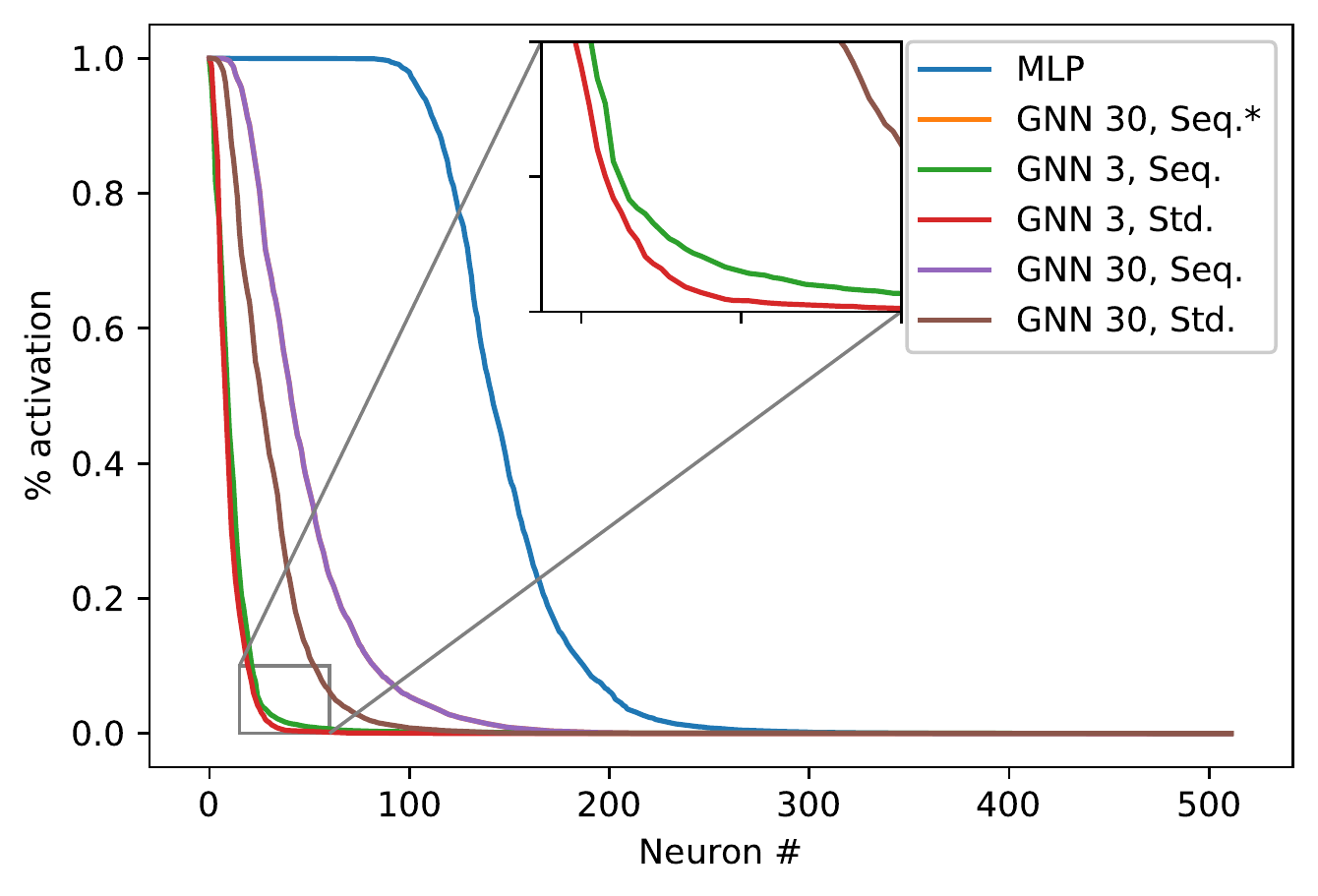} 
    \end{tabular}}
    \caption{Neuron activation rate considering the unimodal architecture.}
  \label{f.activation_audioOnly}
\end{figure}

This section provides an energy-efficiency analysis based on neuronal activation behavior. Such an analysis is fundamental since using such models in real-world applications, such as hearing devices embedded systems, for instance, is constrained to energy issues. It is to be noted that here the energy-efficiency is referred to the number of neurons with zero activity. In hardware (e.g., FPGA) zero signal does not propagate in the network and contributes nothing to the dynamic power consumption because of no switching activity. Figure~\ref{f.activation_audioOnly} describes the average rate of the first hidden layer's neurons activation during the training of the unimodal approach, while Table~\ref{t.activation_audioOnly} provides the area under the curve. Notice that the better values are underlined and some less relevant results were omitted from the plot for better visualization. In this context, one can observe that, even though GNNs and the MLP comprise the same architecture, i.e., two hidden layers with $512$ neurons each, around $100$ neurons in the MLP architecture fired $100\%$ of the time, while the same pattern is observed over a range between $10$ and $20$ neurons concerning the GNNs. Moreover, although the best performing model for the clean audio reconstruction task, i.e., the sequential CCA-GNN with $k=30$, is the most energy-consuming approach between all the CCA-GNN-based methods, it still presents a nominal neuron activation rate if compared to the MLP.

\begin{table}[!htb]
	\caption{Area under the curve considering the unimodal architecture neuron activation rate.}
	\begin{center}
		\resizebox{0.6\textwidth}{!}{
			\begin{tabular}{cccccc}
				\toprule
				Model & Neighbors & Standard & Sequential & Sequential* \\
				\midrule
				\textbf{MLP} & - &  $147.27$ & -& - \\
				\midrule
				\multirow{3}{*}{\textbf{GNN}} & 3 & $\underline{10.17}$ & $11.73$& $11.73$ \\ 
				& 10 &  $21.64$ & $26.56$& $26.56$ \\
				& 30 &  $29.99$ & $47.71$& $47.71$\\ 
				\bottomrule
		\end{tabular}}
		\label{t.activation_audioOnly}
	\end{center}
\end{table}

\begin{figure*}[!htb]
  \centerline{
    \begin{tabular}{cc}
	\includegraphics[width=.44\textwidth]{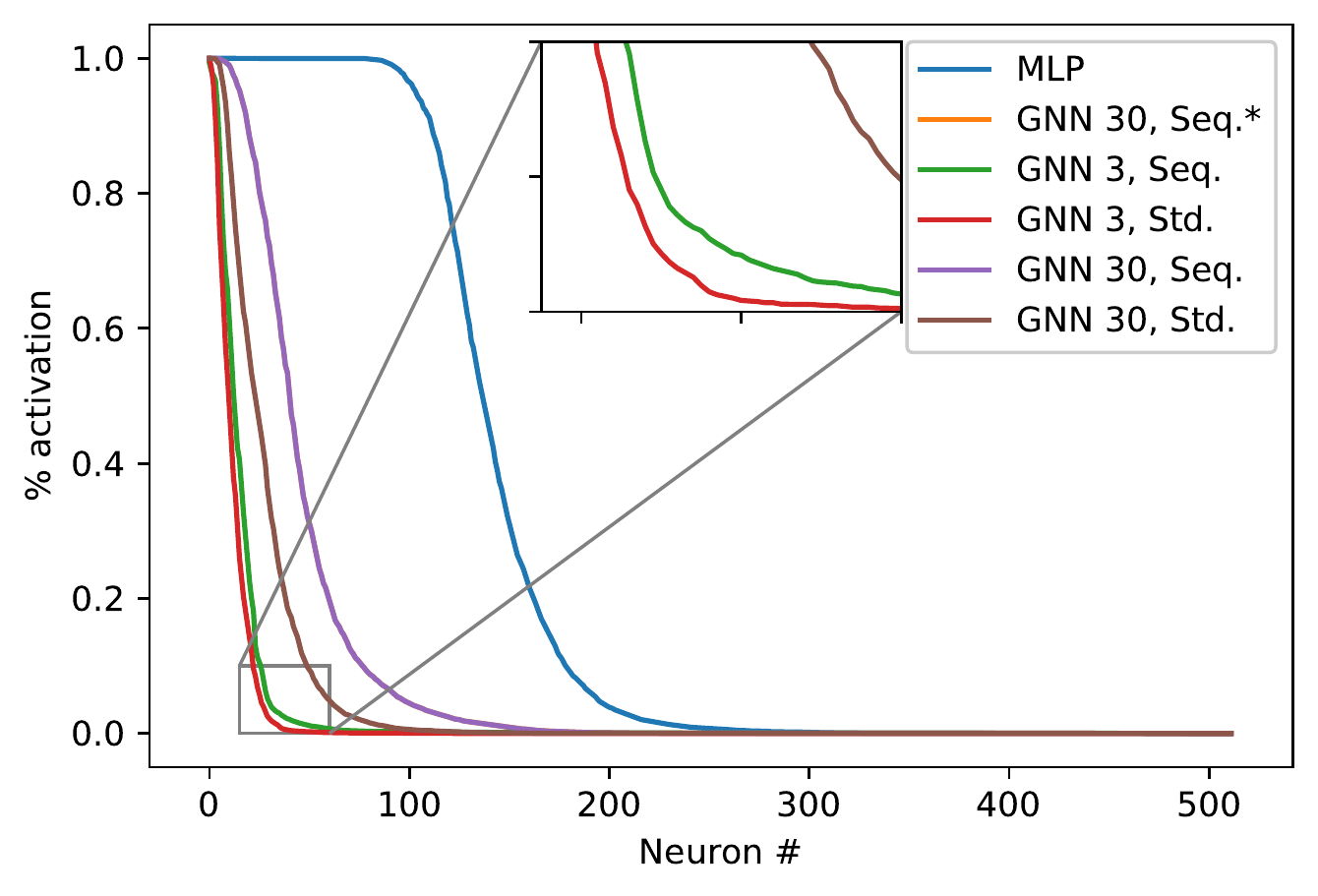} &
	\includegraphics[width=.44\textwidth]{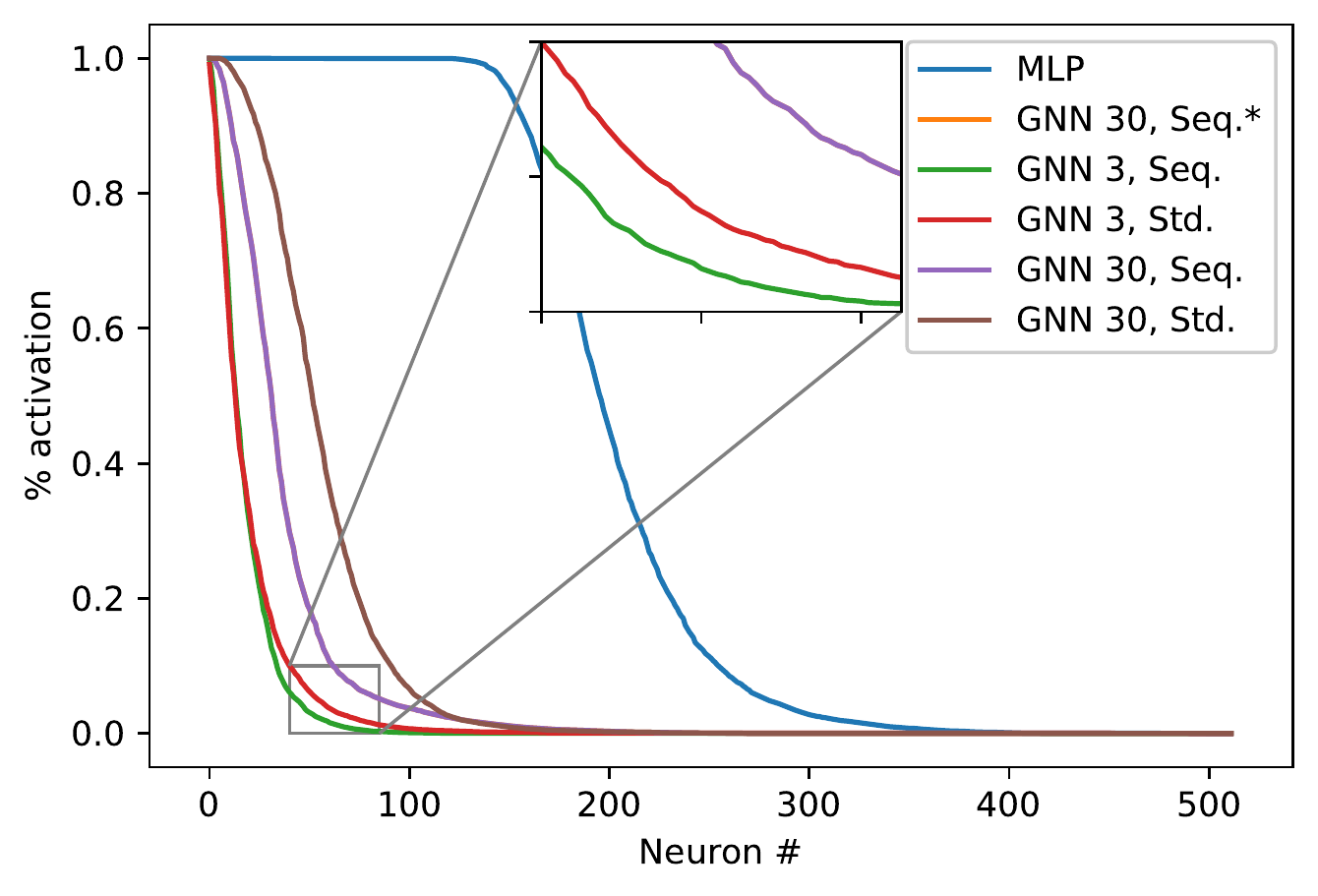} \\
	(a) & (b)
    \end{tabular}}
    \caption{Neuron activation rate considering the multimodal architecture considering (a) audio and (b) visual channels.}
  \label{f.activation_audioVisual}
\end{figure*}

Figures~\ref{f.activation_audioVisual}(a) and~\ref{f.activation_audioVisual}(b) present the neural activation analysis considering the multimodal approach for the audio and the visual channels, respectively, whose values are reflected in Table~\ref{t.activation_audioVisual}. The figures present a similar behavior to the one observed in Figure~\ref{f.activation_audioOnly}, in which the MLP presents a neuron activation rate substantially larger than the MLP-based approaches. Moreover, one can also observe that the prior frame-based approach has different behaviors for each channel since their neuron activation conduct is more intense in the audio channel but produces a reduced fire rate considering the visual channel. Such behavior may suggest that using the prior frame-based approach implies a network inclined to extract more relevant features considering the temporal flow as a context, thus becoming less dependant on the visual context itself.

\begin{table}[!htb]
	\caption{Area under the curve considering the multimodal architecture neuron activation rate.}
	\begin{center}
		\resizebox{0.9\textwidth}{!}{
			\begin{tabular}{ccccccccc}
				\toprule
				&&\multicolumn{3}{c}{\textbf{Audio}}& & \multicolumn{3}{c}{\textbf{Visual}}\\
				\cmidrule{3-5} \cmidrule{7-9} 
				Model & Neighbors & Standard & Sequential & Sequential* & & Standard & Sequential & Sequential* \\
				\midrule
				\textbf{MLP} & - & $142.07$ & -               & - & & $203.01$ & -& -\\
				\midrule
				\multirow{3}{*}{\textbf{GNN}} & 3 & $\underline{11.77}$ & $14.88$& $14.88$  &  & $19.05$ & $\underline{17.36}$& $\underline{17.36}$\\ 
				& 10 &   $19.29$ & $24.01$& $24.01$ & & $38.52$ & $32.28$& $32.28$ \\
				& 30 &  $27.09$ & $45.80$& $45.80$ & & $55.85$ & $36.66$& $36.66$\\ 
				\bottomrule
		\end{tabular}}
		\label{t.activation_audioVisual}
	\end{center}
\end{table}

\subsection{Speech Enhancement Framework and Results} 
\begin{figure}[!htb]
  \centerline{
    \begin{tabular}{c}
	\includegraphics[width=.7\textwidth]{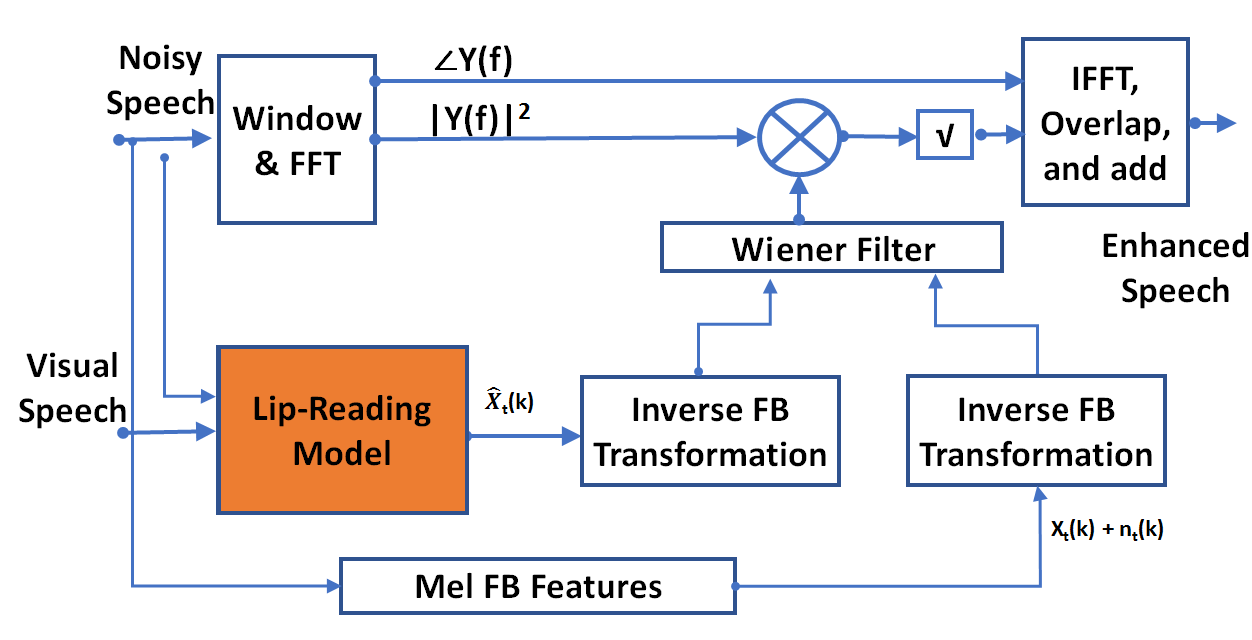} 
    \end{tabular}}
    \caption{Enhanced visually-derived Wiener filtering. Lip reading refers to audio-visual clean log-FB features estimation.}
  \label{f.VWF}
\end{figure}

For speech enhancement, state-of-the-art enhanced visually derived Wiener filter (EVWF) is used as shown in Figure~\ref{f.VWF}~\cite{adeel2019lip}. The EVWF uses an audio or AV driven regression model to estimate clean audio features using noisy audio or AV temporal features. The low dimensional estimated clean audio features are transformed back into high dimensional clean audio power spectrum using inverse filter bank transformation to calculate Wiener filter. The Wiener filter is applied to the magnitude spectrum of the noisy input audio signal, followed by the inverse fast Fourier transform, overlap, and combining processes to produce enhanced magnitude spectrum. More details are comprehensively presented in our previous work \cite{adeel2019lip}\cite{adeel2020contextual}\cite{adeel2020novel}. For objective testing and comparison, perceptual evaluation of speech quality (PESQ) is used as shown in Tables~\ref{t.PESQ_audio} and~\ref{t.PESQ_audioVisual} for audio only and audio-visual data, respectively. PESQ is one of the most reliable methods to evaluate the quality of restored speech. The PESQ score is computed as a linear combination of the average disturbance value and the average asymmetrical disturbance values. Scores range from $-0.5$ to $4.5$, corresponding with low to high speech quality. It can be seen that at low SNR levels (-6dB), EVWF with AV-GNN30* performs comparably to the state-of-the-art LSTM model and significantly outperforms spectral subtraction (SS) and Log-Minimum Mean Square Error (LMMSE), and MLP based speech enhancement methods using far few neurons.  

 \begin{table}[!htb]
\caption{Speech Enhancement Results: Perceptual evaluation of speech quality (PESQ) for the unimodal architecture trained over
noisy audio data considering the standard approach and the proposed prior frame-based positional encoding for time sequence modeling. It is to be noted that only GNN30 and GNN30* with prior frame-based positional encoding perform reasonably well compared to all other models.}
\begin{center}
\renewcommand{\arraystretch}{1.5}
\setlength{\tabcolsep}{6pt}
\resizebox{0.9\textwidth}{!}{
\begin{tabular}{c|c|c|c|c|c|c|c|c|c}
\hhline{-|-|-|-|-|-|-|-|-|-|}
\hhline{-|-|-|-|-|-|-|-|-|-|}
\hhline{-|-|-|-|-|-|-|-|-|-|}
\cellcolor[HTML]{EFEFEF}&\multicolumn{4}{c|}{\cellcolor[HTML]{EFEFEF}\textbf{Standard}}&\multicolumn{5}{c}{\cellcolor[HTML]{EFEFEF}\textbf{Prior frame-based}}\\\hhline{~---------}
\multirow{-2}{*}{\cellcolor[HTML]{EFEFEF}{\textbf{SNR}}}&
  {\cellcolor[HTML]{EFEFEF}{\textbf{MLP}}}& {\cellcolor[HTML]{EFEFEF}{\textbf{GNN 3}}}& {\cellcolor[HTML]{EFEFEF}{\textbf{GNN 10}}}& {\cellcolor[HTML]{EFEFEF}{\textbf{GNN 30}}}&   {\cellcolor[HTML]{EFEFEF}{\textbf{MLP}}}& {\cellcolor[HTML]{EFEFEF}{\textbf{GNN 3}}}& {\cellcolor[HTML]{EFEFEF}{\textbf{GNN 10}}}& {\cellcolor[HTML]{EFEFEF}{\textbf{GNN 30}}}&
  {\cellcolor[HTML]{EFEFEF}{\textbf{GNN 30*}}} \\ \hline
  -12dB &   0.83 &  0.86& 0.80  & 0.88 &  0.80 & 0.91& 0.90 &  1.24 & 1.27\\ \hline
  -6dB &   0.81 &  0.84& 0.85  & 0.87 &  0.82 & 0.83& 0.99 &  1.26 & 1.28\\ \hline
  -3dB &  0.82 &  0.88& 0.89  & 0.94 &  0.86 & 0.87& 0.85 &  1.26 & 1.30\\ \hline
  0dB &   0.85 &  0.85& 0.87& 0.88 &  0.90 & 0.96& 0.88 &  1.33 & 1.34\\ \hline
  3dB &   0.88 &  0.82& 0.91 & 0.86 &  0.86 & 0.92&0.92 &  1.36 & 1.38\\ \hline
  6dB &   0.84 &  0.90& 0.88 & 0.82 &  0.87 & 0.84& 0.89 &  1.40 & 1.42\\ \hline
  12dB &   0.91 &  0.89 & 0.92 & 0.95 &  0.82 & 0.87& 0.81 &  1.48 & 1.55\\ \hline\hline 
\hhline{-|-|-|-|-|-|-|-|-|-|}
\hhline{-|-|-|-|-|-|-|-|-|-|}
\hhline{-|-|-|-|-|-|-|-|-|-|}
\end{tabular}}
\label{t.PESQ_audio}
\end{center}
\end{table}

 \begin{table}[!htb]
\caption{Speech Enhancement Results: Perceptual evaluation of speech quality (PESQ) for the multimodal architecture trained over
noisy audio and clean visual data considering the standard approach and the proposed prior frame-based positional encoding for time sequence modeling. Here only GNN30 without prior frame-based positional encoding and GNN30 and GNN30* with prior frame-based positional encoding perform reasonably well compared to all other models. These models also perform slightly better than audio-only models.}
\begin{center}
\renewcommand{\arraystretch}{1.5}
\setlength{\tabcolsep}{6pt}
\resizebox{\textwidth}{!}{
\begin{tabular}{c|c|c|c|c|c|c|c|c|c|c|c}
\hhline{-|-|-|-|-|-|-|-|-|-|-|-|}
\hhline{-|-|-|-|-|-|-|-|-|-|-|-|}
\hhline{-|-|-|-|-|-|-|-|-|-|-|-|}
\cellcolor[HTML]{EFEFEF}&\multicolumn{4}{c|}{\cellcolor[HTML]{EFEFEF}\textbf{Standard}}&\multicolumn{5}{c|}{\cellcolor[HTML]{EFEFEF}\textbf{Prior frame-based}}&\multicolumn{2}{c}{\cellcolor[HTML]{EFEFEF}\textbf{Obtained from~\cite{adeel2018real}}}\\\hhline{~-----------}
\multirow{-2}{*}{\cellcolor[HTML]{EFEFEF}{\textbf{SNR}}}&
  {\cellcolor[HTML]{EFEFEF}{\textbf{MLP}}}& {\cellcolor[HTML]{EFEFEF}{\textbf{GNN 3}}}& {\cellcolor[HTML]{EFEFEF}{\textbf{GNN 10}}}& {\cellcolor[HTML]{EFEFEF}{\textbf{GNN 30}}}&   {\cellcolor[HTML]{EFEFEF}{\textbf{MLP}}}& {\cellcolor[HTML]{EFEFEF}{\textbf{GNN 3}}}& {\cellcolor[HTML]{EFEFEF}{\textbf{GNN 10}}}& {\cellcolor[HTML]{EFEFEF}{\textbf{GNN 30}}}&
  {\cellcolor[HTML]{EFEFEF}{\textbf{GNN 30*}}}& {\cellcolor[HTML]{EFEFEF}{\textbf{SS}}}&
  {\cellcolor[HTML]{EFEFEF}{\textbf{LMMSE}}} \\ \hline
  -12dB &   0.84 &  0.88& 0.87  & 1.10 &  0.86 & 0.94& 0.80 &  1.25 & 1.29 & 0.9& 0.95\\ \hline
  -6dB &   0.88 &  0.80& 0.86  & 1.15 &  0.91 & 0.80& 0.89 &  1.27 & 1.29& 1.01 &1.03\\ \hline
  -3dB &  0.90 &  0.87& 0.89  & 1.15 &  0.83 & 0.86& 0.88 &  1.28 & 1.31 & 1.17 &1.18\\ \hline
  0dB &   0.82 &  0.86& 0.81& 1.18 &  0.80 & 0.84& 0.84 &  1.34 & 1.36 & 1.21& 1.20\\ \hline
  3dB &   0.86 &  0.89& 0.93 & 1.20 &  0.92& 0.88&0.88 &  1.40 & 1.42 & 1.25 &1.34\\ \hline
  6dB &   0.89 &  0.87& 0.88 & 1.22 &  0.87 & 0.83& 0.81 &  1.44 & 1.48 & 1.26 & 1.39\\ \hline
  12dB &   0.84 &  0.91 & 0.84 & 1.25 &  0.88 & 0.81& 0.89 &  1.51 & 1.60 &1.54 &1.60\\
\hhline{-|-|-|-|-|-|-|-|-|-|-|-|}
\hhline{-|-|-|-|-|-|-|-|-|-|-|-|}
\hhline{-|-|-|-|-|-|-|-|-|-|-|-|}
\end{tabular}}
\label{t.PESQ_audioVisual}
\end{center}
\end{table}

\section{Conclusions}
\label{s.conclusions}

This paper proposes an energy-efficient approach for improving and boosting sound signals through environmental information fusion. The model extends Graph Neural Networks with Canonical Correlation Analysis for multimodal data integration and further incorporates a prior frame-based node's positional encoding that considers the temporal sequence in data to establish the information similarity instead of the usual feature space distance. Experiments conducted over the AV Grid and ChiME3 corpora considering the task of clean audio reconstruction based on the fusion of noisy audio and clean video data show that the proposed approaches are capable of outperforming a baseline composed of an MLP/ LSTM with similar architecture trained under the same conditions, i.e., in a self-supervised fashion using canonical correlation analysis maximization as the target function. Finally, the proposed approach provides a considerable gain regarding energy efficiency, given that the CCA-GNN neuron firing rates are dramatically lower than MLP and LSTM. It is worth mentioning that it is impossible to quantify the energy saving at this stage, but the ongoing work includes implementing these models on FPGA in which neurons with zero activity will not propagate in the network and will therefore contribute nothing to the dynamic power consumption. 

The experiments also indicate that the multimodal approach can produce better results than the unimodal architecture, leading to minor reconstruction errors. Such behavior is expected since the visual data acts as a context to introduce some additional meaning and improve the noisy audio signal. Additionally, the prior frame-based approach provided better results than the standard model, showing the importance of the temporal information as an additional context to the noisy signal. Finally, one could notice by the neuron's firing behavior that using prior frame node connection reinforces the information present in the noisy audio channel, making it less dependant on the visual data context for clean audio reconstruction. Future work involves developing a more biologically realistic neuronal model, introducing a concept of memory to improve communication mechanisms between the channels. Further, Graph Neural Networks with Canonical Correlation Analysis will be used to improve cross channels communication blocks within convolutional neural networks. 

\section*{Acknowledgments}
\begin{sloppypar}
This research was supported by the UK Engineering and Physical Sciences Research Council (EPSRC) Grant Ref. EP/T021063/1. J. Del Ser would like to thank the Spanish Centro para el Desarrollo Tecnologico Industrial (CDTI, Ministry of Science and Innovation) through the ``Red Cervera'' Programme (AI4ES project), as well as by the Basque Government through EMAITEK and ELKARTEK (ref. 3KIA) funding grants and the Consolidated Research Group MATHMODE (ref. IT1456-22).
\end{sloppypar}

\section*{References}

\bibliography{refs}

\end{document}